\documentclass{aa}
\usepackage{graphicx}

\usepackage{siunitx}

\usepackage[flushleft]{threeparttable}

\usepackage{subcaption}

\usepackage{txfonts}

\usepackage{natbib,twoopt}

\usepackage[colorlinks=true, linkcolor=blue, citecolor=blue, urlcolor=blue, breaklinks=true]{hyperref} 
\bibpunct{(}{)}{;}{a}{}{,}           
\makeatletter
  \newcommandtwoopt{\citeads}[3][][]{\href{http://adsabs.harvard.edu/abs/#3}%
    {\def\hyper@linkstart##1##2{}%
     \let\hyper@linkend\@empty\citealp[#1][#2]{#3}}}
  \newcommandtwoopt{\citepads}[3][][]{\href{http://adsabs.harvard.edu/abs/#3}%
    {\def\hyper@linkstart##1##2{}%
     \let\hyper@linkend\@empty\citep[#1][#2]{#3}}}
  \newcommandtwoopt{\citetads}[3][][]{\href{http://adsabs.harvard.edu/abs/#3}%
    {\def\hyper@linkstart##1##2{}%
     \let\hyper@linkend\@empty\citet[#1][#2]{#3}}}
  \newcommandtwoopt{\citeyearads}[3][][]%
    {\href{http://adsabs.harvard.edu/abs/#3}
    {\def\hyper@linkstart##1##2{}%
     \let\hyper@linkend\@empty\citeyear[#1][#2]{#3}}}
\makeatother

\usepackage{color}

\usepackage{chemist}

\usepackage{orcidlink}

\begin{document}

\title{Reconstruction of Cepheid Radial Velocity Curves from the shape of the V-band Light Curves}

\titlerunning{Reconstruction of Cepheid Radial Velocity Curves}
\authorrunning{Hocd\'e et al. }

\author{V.~Hocd\'e\inst{1}\orcidlink{0000-0002-3643-0366}
\and P.~Moskalik\inst{2}\orcidlink{0000-0003-3142-0350}
\and N.~Nardetto\inst{1}\orcidlink{0000-0002-7399-0231}
\and P.~Kervella\inst{3,4}
\and B.~Pilecki\inst{2}
\and R.~Smolec\inst{2}
\and G.~Pietrzyński\inst{2}
\and W.~Gieren\inst{6}
\and G.~Hajdu\inst{2}
\and A.~Gallenne\inst{5}
\and M.~C.~Bailleul\inst{1}
\and G.~Bras\inst{3}
\and P. Wielgórski\inst{2}
\and L.~Breuval\inst{7}
\and A.~Mérand\inst{8}
\and R.~S.~Rathour\inst{1}
\and F.~Espinoza-Arancibia\inst{2}
\and W.~Kiviaho\inst{3}
\and B.~Apostolova\inst{1} 
\and K.~Sivkova\inst{3} }

\institute{Universit\'e Côte d'Azur, Observatoire de la C\^ote d'Azur, CNRS, Laboratoire Lagrange, France\\
email : \texttt{vincent.hocde@oca.eu}
   \and 
   Nicolaus Copernicus Astronomical Centre, Polish Academy of Sciences, Bartycka 18, 00-716 Warszawa, Poland
   \and LIRA, Observatoire de Paris, Université PSL, Sorbonne Université, Université Paris Cité, CY Cergy Paris Université, CNRS, 5
place Jules Janssen, 92195 Meudon, France
\and French-Chilean Laboratory for Astronomy, IRL 3386, CNRS, Casilla 36-D, Santiago, Chile
\and Instituto de Alta Investigación, Universidad de Tarapacá, Casilla 7D, Arica, Chile
\and Universidad de Concepción, Departamento Astronomía, Casilla 160-C, Concepción, Chile
\and European Space Agency (ESA), ESA Office, Space Telescope Science Institute, 3700 San Martin Drive, Baltimore, MD 21218, USA
\and European Southern Observatory Headquarters, Karl-Schwarzschild-Str. 2, 85748 Garching, Germany
}

\date{Received ... ; accepted ...}

\abstract
{Radial velocity (RV) curves of Cepheids are essential for studying their pulsation properties and also for measuring their distance via the parallax-of-pulsation method. Although precise RV curves are available for hundreds of Cepheids, it is still not possible to predict the complete RV curve from other observables such as their pulsation period or metallicity.}
{This paper aims to develop the first method to reconstruct the shape of the RV curves of short-period fundamental-mode Cepheids, based exclusively on their pulsation period and the morphology of their $V$-band light curves (LCs).}
{We compiled a dataset of high-quality spectroscopic and photometric measurements from the literature for 81 short-period fundamental-mode Galactic Cepheids up to a pulsation period of 8\,days, enabling precise determination of the Fourier parameters and their uncertainties. By performing a detailed comparative analysis of their shapes, we investigated correlations between LC and RV Fourier parameters and used these relations to reconstruct the RV curves. We further assessed the accuracy of these reconstructions by examining potential metallicity effects with an additional dataset of 23 metal-poor Cepheids.}
{For pulsation periods between 3.5 and 7.0\,days, we found
tight correlations between different combinations of LC and RV
Fourier parameters up to order 7. In particular, we discovered that the ratios $R_{21}(RV)/R_{21}(LC)$ and $R_{31}(RV)/R_{31}(LC)$ are tightly correlated with the pulsation period. These relationships enable the reconstruction of RV curves of Cepheids with their LC. The reconstructed curve has an uncertainty of about 0.60${\rm km\,s}^{-1}$ relative to the Fourier fit of true spectroscopic RV measurements. For individual Cepheids, the reconstructed RV curves integrated along the pulsation cycle (i.e.
the linear radius variations) are accurate to less than 1\% and precise to within 4.16\% in comparison to the integrated true spectroscopic RV curves. Last, our sample of metal-poor Cepheids shows a good agreement with the empirical relations calibrated on Galactic pulsators, indicating that the method of reconstruction is weakly dependent on metallicity.}
{For the first time, we present a method to reconstruct the shape of the RV curves of Cepheids using solely the $V$-band LC and the pulsation period. This approach provides a valuable tool for the reconstruction of RV curves for extragalactic Cepheids through photometric data alone. It opens the road to a purely photometric parallax-of-pulsation method in the context of photometric surveys, such as the Vera Rubin Telescope. Further calibration in different photometric bands and for a larger metallicity baseline will be useful to improve the method.}
{}

\maketitle

\section{Introduction}\label{Intro}
The radial velocity (RV) curves of Cepheids are essential to measure the radius variation in the application of the Baade-Wesselink (BW) method  \citep{lindermann18,baade26,VanHoof1945,wesselink46} for distance measurement. This technique requires high-resolution spectroscopic observations of metallic lines for accurate RV measurements \citep{borgniet2019}, which is currently challenging to obtain for distant Cepheids beyond the Galaxy. Up to now, only a few dozens of RV curves have been obtained for Cepheids in the Large and Small Magellanic Clouds (LMC and SMC) for the application of the BW method \citep{storm11b,Gieren2018}, mostly for the brightest, long-period Cepheids, and for which the pulsation cycle is sparsely covered. The upcoming Gaia DR4 will mark a major step forward, providing RV measurements for thousands of Cepheids in the Magellanic Clouds from Calcium Triplet observations with the Radial Velocity Spectrometer (RVS) \citep{gaia2023,Katz2023}. Despite these efforts, spectroscopic observations of extragalactic Cepheids within the Local Group remain technically challenging. To date, only 2 RV curves of Cepheids were presented in a conference article, respectively in M31 and M33 \citep{Forestell2004}. For the future Extremely Large Telescope (ELT) in combination with multi-fibers instruments \citep{ANDES2024,MOSAIC2024}, such observations will be limited to the brightest targets and possibly a sparse coverage of the pulsation cycle.

On the other hand, RV curves of Cepheids are available for more than two hundreds Cepheids in the solar neighborhood \citep[][and references therein]{Anderson2024,Hocde2024RV}.
The Fourier decompositions of these RV curves improved our understanding of their pulsation, and are also useful for various purposes, including comparison with \textit{Gaia} RVS data \citep{Anderson2024}, constraining pulsation models \citep{Paxton2019,Molinaro2025}, disentangling pulsation and orbital motion \citep{Nardetto2024,Shetye2024}, and pulsation mode identification \citep{Kienzle1999,Hocde2024RV}. The analysis of Cepheid RV curves through their Fourier parameters reveals a diversity of shapes and amplitudes that likely depend on the physical parameters of each star, similarly to what is observed for their LC morphologies \citep{Bhardwaj2015,Hocde2024RV}.

Despite this complexity, LCs and RV curves of fundamental-mode Cepheids depend primarily on the pulsation period and their shape follows the so-called \textit{Hertzsprung} progression \citep{Ludendorff1919,Hertzsprung1926}. This shape progression was observed among LCs of Cepheids in the Magellanic Clouds and Andromeda Galaxy \citep{Parenago1936,Shapley1940,Payne-Gaposchkin1947} and later confirmed for the RV curves \citep{Joy1937,LedouxWalraven1958}. This progression is efficiently described by Fourier decomposition of the curves \citep{SimonLee1981,1983SimonTeays} which can then be compared to hydrodynamical models \citep{Buchler1990,Moskalik1992}. The Hertzsprung progression is explained
theoretically by the resonance between the fundamental and the second overtone pulsation modes characterized by pulsation periods of $P_0$ and $P_2,$ respectively, as suggested first by \cite{SimonSchmidt1976} and then demonstrated analytically with the amplitude equation formalism \citep{Buchler1986,kovacs1989}.
In particular, the Fourier phase difference $\phi_{21}$ of the RV curve of the fundamental-mode depends almost exclusively on the resonant period ratio $P_2$/$P_0$. The progression of this Fourier phase, both for
the RV curves and for the LCs, is characterized by a pronounced discontinuity at the resonance center at about P = 10\,days. As the resonance is a dynamical phenomenon, RV curves provides information on the stellar structure and dynamics of Cepheids.  On the other hand, LCs of Cepheids depend not only on the pulsation dynamics, but also on the radiative transfer through the outer layers of the photosphere.

Comparison of RV curves and LCs are rare in the literature because it demands a significant sample of stars with both well sampled LC and RV curves. To our knowledge, only the amplitude ratio $A_{RV}/A_V$ was studied theoretically and observationally by \cite{balona79b} and \cite{KlagyivikSzabados2009}, respectively, in the objective of better discriminate the pulsation mode of Cepheids.  Understanding the link between LC and RV curves can be useful for constraining hydrodynamical pulsation models, but also to reconstruct the RV curves of Cepheids from the shape of the LCs. The objective of this paper is to propose a method for reconstructing the full RV curves of Cepheids on the basis of the $V$-band light curves. This new technique could be useful to complement spectroscopic observations to recover full RV curves. In the long term, the method could be combined with large-scale photometric surveys to apply the BW technique to more distant Cepheids in the Local Group, ultimately improving the calibration of the cosmic distance scale.  This is particularly relevant in the framework of photometric surveys such as Gaia DR4 or the Vera C. Rubin Observatory which will obtain LCs to thousands of Cepheids \citep{Hoffman2015}.

In the following we first present the calibrating sample in Sect.~\ref{sect:calibration} and the calculation of Fourier parameters of LCs and RV curves in Sect.~\ref{sect:fourier}. We then use these parameters to fit several empirical relations between LC and RV curves in Sect.~\ref{sect:templates}. In Sect.~\ref{sect:uncertainties}, we quantify the internal and external accuracy of the reconstructed shape of the RV curves, and we quantify the accuracy of the template fitting method in Sect.~\ref{sect:template_fitting}. We discuss future improvements of the method in Sect.~\ref{sect:discussion} and we conclude in Sect.~\ref{sect:conclusion}.

\section{Calibrating sample}\label{sect:calibration}

\subsection{Choice of period range}
In this paper, we focus on fundamental-mode Cepheids between 3 and 8\,days (logP$\approx$0.48 to 0.91). In the Galaxy, this is considered to be a short-period range, while in metal-poor environments such as the LMC it might be considered as an intermediate-period range. As we will calibrate our method on Galactic Cepheids we will refer to short-period Cepheids in the following. The choice of this narrow period range simplifies several aspects compared to their long-period counter-parts. First, the Fourier parameters, particularly the Fourier phases of RV curves exhibit tighter and more regular progressions in this period range. This makes these stars ideal to construct a predictive method. Second, the population of Cepheids with well sampled LC and RV curves is also larger for short-period Cepheids compared to their long-period counterparts, which also allows a more robust study of their Fourier parameters.

Another advantage of using short-period Cepheids is due to their slower evolutionary period change of the order of 0.1 s/yr during their second and third crossing of the instability strip \citep{Turner2006,Csornyei2022} which largely mitigates the epoch difference between RV curves and LCs in comparison to fast evolving long-period Cepheids. We note however that some short-period Cepheids might undergo a fast period change because they are crossing the instability strip for the first time \citep{Rodriguez2022,Espinoza2022}, but such objects are rare. Anomalously large rates of period change are also found in several
Galactic Cepheids with pulsation periods below 3.5 days \citep{Turner2006}. Again, such objects are very rare and they are not represented in
our calibrating sample.

Last, the 2:1 resonance at about $P=10\,$days also introduces discontinuity in the RV and LC Fourier progression which are more difficult to model empirically. Therefore we limit our sample to a pulsation period of 8\,days. At the lower end of the pulsation period, fundamental-mode Cepheids pulsating in less than 3 days are rare in the Galaxy, contrary to what is observed for LMC and SMC Cepheids \citep{Soszynski2008,Soszynski2010}. In the following, we build a calibrating sample for short-period Cepheids between 3 and 8\,days.

\subsection{Source of RV curves}\label{sect:Source_RV}
In this work we gather the most precise Fourier parameters for RV curves of fundamental-mode short-period Cepheids. In our recent work we provided precise Fourier parameters for 72 fundamental-mode short-period Cepheids RV curves \citep{Hocde2024RV}. At the same time, new precise RV measurements were made available for 64 short-period fundamental-mode Cepheids \citep{Anderson2024}. From these two datasets, we used the best quality fits yielding the most precise Fourier parameters. To ensure consistency in the evaluation of RV curve quality and derivation of the uncertainties, we re-computed the Fourier parameters from \citet{Anderson2024} using the method described in \citet{Hocde2024RV} (see also Sect.~\ref{sect:fourier}). From these two samples, we selected only the stars with the best fits, based on the smallest dispersion of the fit and the absence of significant fit instabilities \citep[labeled Quality 1 and 1a,][]{Hocde2024RV}. These requirements ensure the accuracy and the precision of the Fourier parameters.

In order to complement the sample with metal-poor Cepheids, we relaxed these constraints for additional RV data. Hence, we gathered RV curves of metal-poor Cepheids observed in the LMC \citep{storm11b,Pilecki2018} and the SMC \citep{Gieren2018}. We also gathered Fourier parameters derived by \cite{Hocde2024RV} from metal-poor Cepheids observed by \cite{Pont2001} in the outskirt of the Milky Way at a galactocentric radius of about 12-14\,kpc. We include metal-poor Cepheids with pulsation period down to 2\,days. This sample of stars allows the investigate of potential metallicity effect in the LMC and SMC range. However, most of the RV measurements of these stars suffer from several limitations, including fewer number of measurements, sparse coverage of the pulsation cycle and intrinsic scatter, both of which lead to significant instabilities in the Fourier fits. The only exceptions are the spectroscopic observations from \cite{pilecki13,Pilecki2018} for two LMC Cepheids with good quality. Due to the overall poor quality of Fourier fits for metal-poor Cepheids, we decide to not include them in the calibration and used only to test consistency with the MW calibrating sample. 

From this sample of RV curves we selected only fundamental-mode Cepheids based on a robust pulsation mode identification using $\phi_{21}$ from the RV curve \citep{Kienzle1999,Hocde2024RV}. This is particularly important for Cepheids with pulsation period above 5.5 days, where other selection criteria based on the LC shapes are ambiguous.

Our sample of RV curves consist of 81 fundamental-mode Cepheids of pulsation period between 3.45 and 7.99 days (see Table \ref{tab:data_ref}), with a median fit dispersion of $\sigma_\mathrm{fit}=0.5\,$km/s. In a separate sample, not used for the calibration, we include 23 RV curves for metal-poor stars with pulsation period between 2.64 and 7.46\,days also listed in Table \ref{tab:data_ref_metal_poor}. In the following, we cross-match these samples with LCs sources available in the literature.

\subsection{Source of V-band light curves}

The choice of the $V$-band LCs is motivated by several considerations. First of all, $V$-band LCs are available for a large number of stars which is convenient to cross-match with RV curves. $V$-band LCs are also well sampled along the pulsation cycle, which is necessary for the derivation of precise Fourier parameters. Nevertheless, extending the analysis to other photometric bands will be interesting for future work.
We gathered LCs in the $V$-band from different sources in the literature, as presented in Table \ref{tab:data_ref}. We retrieved most of the LCs sources from the compilation of \cite{Berdnikov2008}, the ASAS catalogue \citep{Pojmanski2002} and ASAS-SN\footnote{All-Sky Automated Survey for Supernovae catalog}\citep{ASAS2018}, KISS\footnote{Kiso Supernova Survey} \citep{Morokuma2014}, SWASP\footnote{Super Wide Angle
Search for Planets} \citep{Pollacco2006}.  In few cases, we complemented the data with older datasets from \cite{moffett84}, \cite{kiss98}. For several stars we combined these data sets to derive more precise Fourier parameters in the next section. For Cepheids of the LMC and SMC  we used mostly the LCs
from the OGLE\footnote{Optical Gravitational
Lensing Experiment} III \citep{Soszynski2008,Soszynski2010} and OGLE IV \citep{Soszynski2015}.

In this sample, several of the Galactic Cepheids are known binaries or member of multiple systems \citep{KlagyivikSzabados2009,Evans2024,Shetye2024}. In most cases the companions are main sequence stars that are much fainter than the Cepheids \citep{Gallenne2019}. To date, V1334 Cyg is the Galactic Cepheid\footnote{V1334 Cyg is an overtone pulsator and is not included in our sample.}  with the brightest detected companion with a flux ratio close to ~10\% in the $V$-band \citep{Evans1995}.  However, other Cepheids, such as OGLE-LMC-CEP-0227, are known to have bright giant companions  \citep{Pilecki2021}. For this particular star, we subtracted the light of the companion \citep[$V=15.90\pm0.05$,][]{Pilecki2018} in order to correct the shape of the $V$-band LC for this star.

\begin{figure}[htbp]
    \centering
    \begin{subfigure}[b]{0.24\textwidth}
        \includegraphics[width=\linewidth]{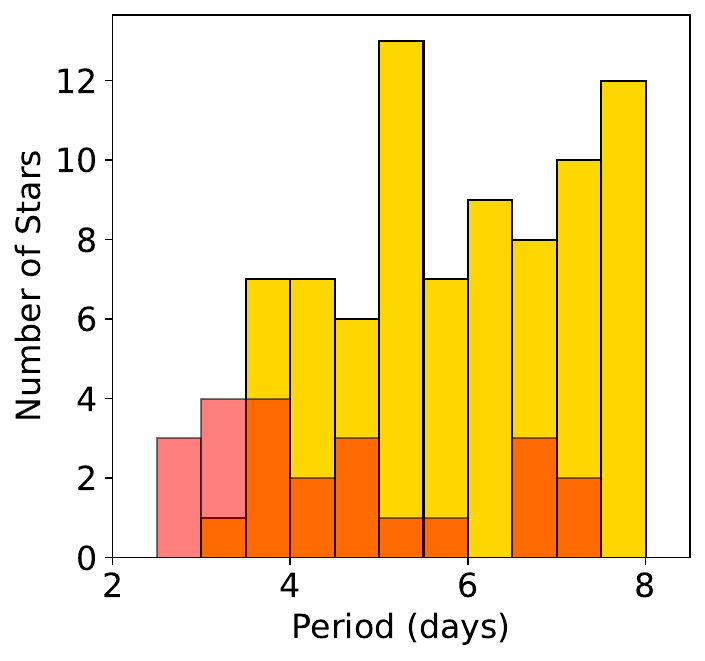}
        \caption{}
        \label{fig:period_histo}
    \end{subfigure}
    \begin{subfigure}[b]{0.24\textwidth}
        \includegraphics[width=\linewidth]{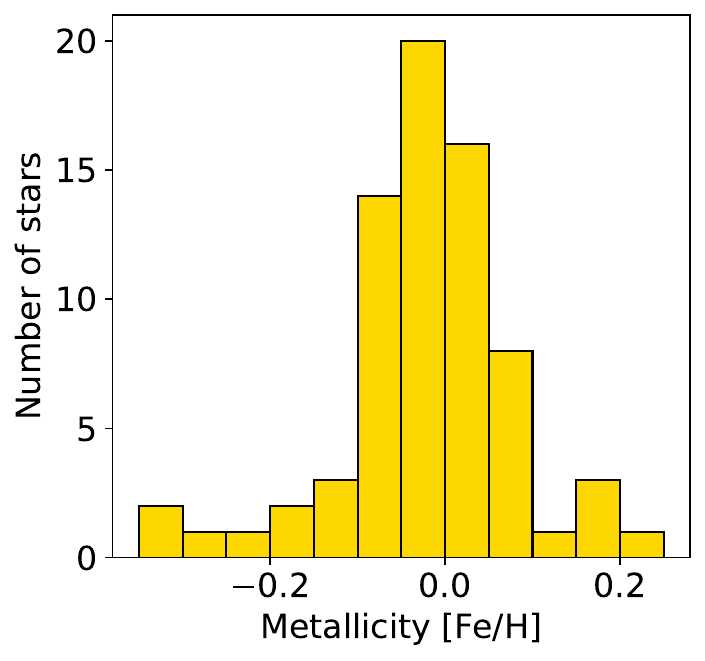}
        \caption{}
        \label{fig:metal_histo}
    \end{subfigure}
    \caption{(a) Histogram of the pulsation period of the calibrating
   sample in yellow and of the metal-poor sample in red. (b) Histogram
   of the metallicity of the calibrating sample only  (see Sect.~\ref{sect:calibration}).}
\end{figure}

Our final calibrating sample comprises 81 Cepheids  with both LCs and RV curves, uniformly distributed between 3.45 and 8\,days, as shown in Fig.~\ref{fig:period_histo}, ensuring a consistent calibration across this period range. This sample is complemented with 23 metal-poor Cepheids from the LMC, SMC and the outskirt of the Galaxy, which are used to verify the effect of metallicity on the empirical calibration. We cross-matched the calibrating sample with the metallicity compilation by \citet{Hocde2023} and identified spectroscopic metallicities for 72 stars, the majority of which are from the determinations of \citet{Luck2018}. We present the histogram of the metallicities of the calibrating sample in Fig.~\ref{fig:metal_histo}. As expected, our calibrating sample is well centered on solar metallicity. On the other hand, the metal-poor samples (not plotted) are well centered on  [Fe/H]=$-0.75$\,dex \citep{Romaniello2008} for SMC stars and $-0.41$ for LMC stars \citep{Romaniello2022}, and subsolar values between $-0.4$-$-0.1$ for Cepheids in the outskirt of the Galaxy \citep{Pont2001,Trentin2024}.

\section{Fourier parameters of LC and RV curves}\label{sect:fourier}
\subsection{Fourier decomposition}
For both the LC and RV curves we applied a Fourier fitting using the least-square method, and computing uncertainties on the Fourier parameters following the method described by \cite{Petersen1986}. In the following, we summarized the steps of the method presented in \cite{Hocde2024RV}. The idea of applying Fourier expansion to periodic LCs was introduced by \cite{Schaltenbrand1971} and further developed by \cite{SimonLee1981}. For each curves $C(t)$ measured at time $t$, we derived the following Fourier decomposition

\begin{equation}\label{eq:fourier}
C(t) = A_0 + \sum_{k=1}^n A_k \mathrm{sin}[k \omega t +\phi_k]
,\end{equation}
where $A_k$ and $\phi_k$ are the amplitude and phase of the $k$ Fourier component. Each harmonic frequency is a multiple of the fundamental frequency $\omega=2\pi/P$. The pulsation period, $P$, is also optimized in the fitting process. The standard deviation of the fit is given by
\begin{equation}\label{eq:sigma}
\sigma^2=\frac{1}{N-2n-1}\sum^N_{i=1}(C(t_i)-\hat{C}_i)^2,
\end{equation}
where $N$ is the number of data points, $n$ the order of the fit, and $C(t_i)$ and $\hat{C}_i$ are measurements and the Fourier model, respectively.  We used the standard deviation of the fit to perform 3$\sigma$-clipping on the Fourier fit to remove the outliers.

\begin{figure}[htbp]
    \centering

    \begin{subfigure}[b]{0.40\textwidth}
        \centering
        \includegraphics[width=\linewidth]{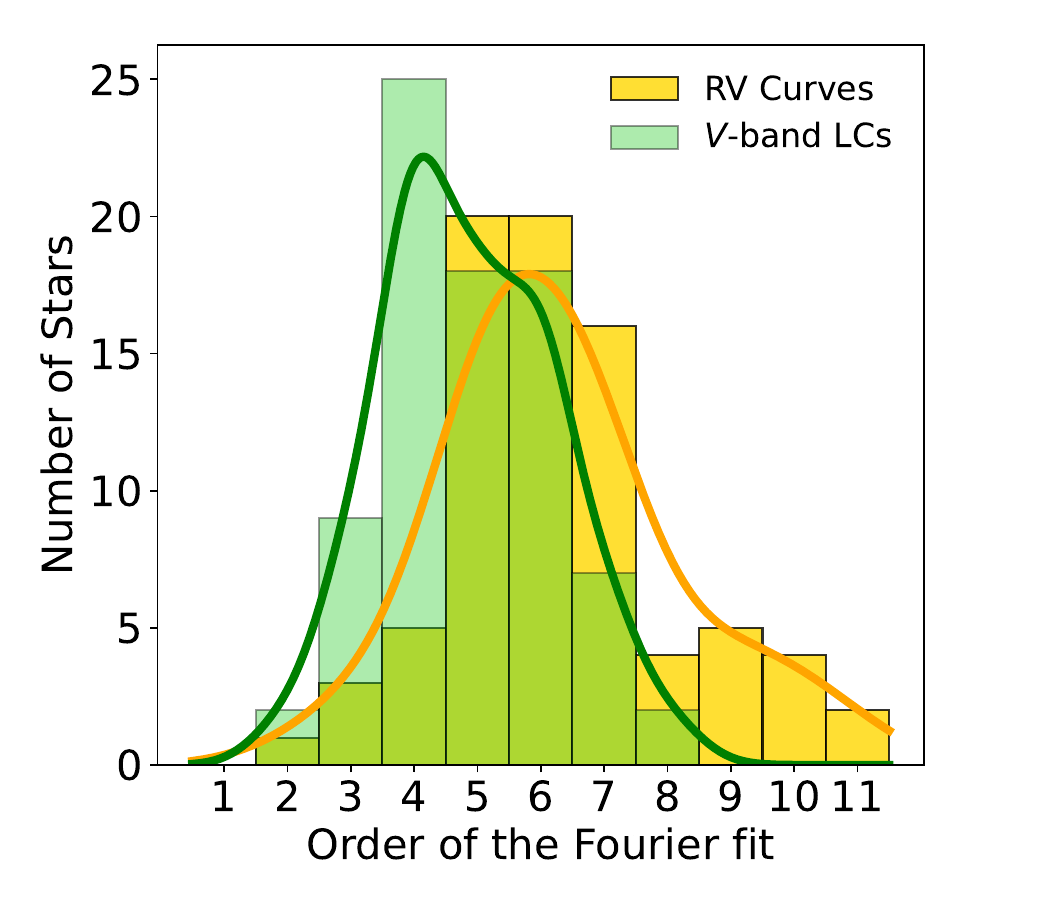}
        \caption{}
        \label{fig:order_histo}
    \end{subfigure}
    \hfill
    \begin{subfigure}[b]{0.42\textwidth}
        \centering
        \includegraphics[width=\linewidth]{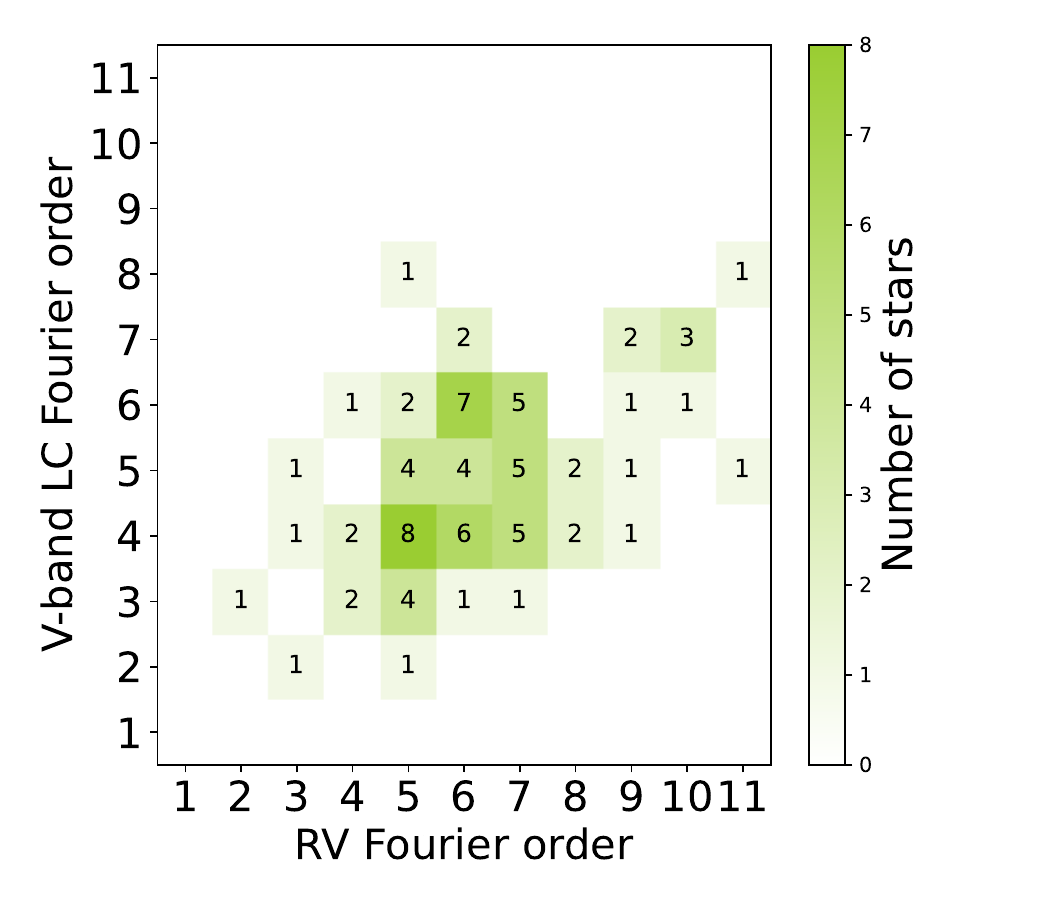}
        \caption{}
        \label{fig:2d_histo}
    \end{subfigure}

    \caption{ (a) Histogram of the order of the Fourier fits for the $V$-band LCs and RV curves of the initial calibrating sample (81 Cepheids up to a pulsation period of 8\,days). (b) 2D histogram of the LC Fourier order versus RV Fourier order.}
    \label{fig:fit_order_global}
\end{figure}

To determine the order $n$ of the fit we adopt an adaptive approach, iteratively increasing $n$ as long as condition $A_n / \sigma_{A_n} > 4$ is satisfied\footnote{In case where Fourier amplitudes do not decrease monotonically, most often for long-period Cepheids (P>10\,d), we include higher significant orders.}. In some cases, one or two additional harmonics were included to stabilize the fit. The histogram of the Fourier order of RV and the LC fits for the calibrating sample is presented in Fig.~\ref{fig:order_histo}, which shows that RV curves require a larger number of harmonics to be described in our sample, with a median order of the fit $n=6$. 

For RV curves we remove possible orbital trends with secondary Fourier sum. To that effect we model the orbital radial velocity with the Fourier sum with the orbital period, $P_\mathrm{orb}$. This method removes orbital motion as our objective is only to model the pulsational component. The only exception are for orbital motion in two LMC Cepheids, which was modeled and removed by \cite{Pilecki2018}. We also correct for possible slow linear or parabolic trends observed in residuals of LCs or RV curves, possibly due to instrumental effects or to orbital motion in the latter case. To this end we model the trend with a single long-period sine component with $P_\mathrm{trend}=50000$\,days. We indicate when we correct for such a trend in the Table \ref{tab:fourier_rv}.

Finally, we used the dimensionless Fourier parameters introduced by \cite{SimonLee1981}: the amplitude ratios of order~$k$
\begin{equation}
    R_{k1}=\frac{A_k}{A_1},
\end{equation}
and the Fourier phase difference, which locates each harmonic with respect to the first harmonic
\begin{equation}
    \phi_{k1}=\phi_k - k \phi_1.
\end{equation}
We present the Fourier parameters and their uncertainties obtained for both LC and RV curves in Tables \ref{tab:fourier_lc} and \ref{tab:fourier_rv}, and we plot them, up to order 6, in Figs.~\ref{fig:fourier_all_RV} and \ref{fig:fourier_all_LC}.

\subsection{Effect of the pulsation period on Fourier parameters}
We note that the Fourier fits yield slightly different pulsation periods for the RV curves and LCs. The difference is always smaller than 0.0003\,day (about 26 seconds), as shown in Table~\ref{tab:data_ref}. This discrepancy may arise from a genuine period change, since the RV and photometric data are not necessarily contemporaneous, or from numerical uncertainties in the fitting procedure. Although small, this difference can still exceed the formal uncertainty of the pulsation period. We verified that phasing of the RV data with the pulsation period derived from the LC (instead of the one derived from the RV) has only a minor effect on the RV Fourier parameters. Even for the stars exhibiting the largest difference between the two periods, the changes of velocity $A_1$ and $R_{n1}$ never exceed the 2$\sigma$ level and usually are smaller. We therefore assume in the following that the pulsation periods of the LC and RV curves are equal and we adopt the RV-derived periods for all future computations.

\subsection{Qualitative description of the Fourier parameters}\label{sect:fourier_discuss}
\begin{figure*}[htbp]
    \centering

    \begin{subfigure}[b]{0.47\textwidth}
        \centering
        \includegraphics[width=\textwidth]{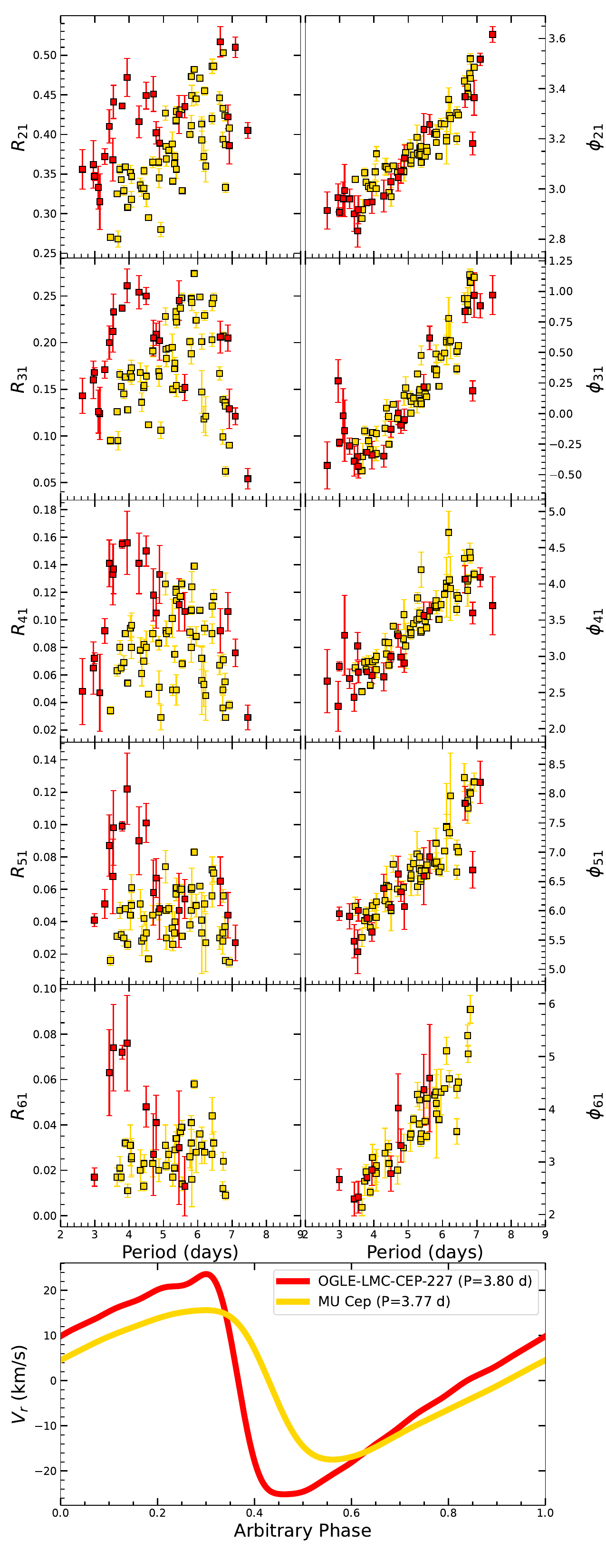}
        \caption{}
        \label{fig:fourier_all_RV}
    \end{subfigure}
    \hfill
    \begin{subfigure}[b]{0.47\textwidth}
        \centering
        \includegraphics[width=\textwidth]{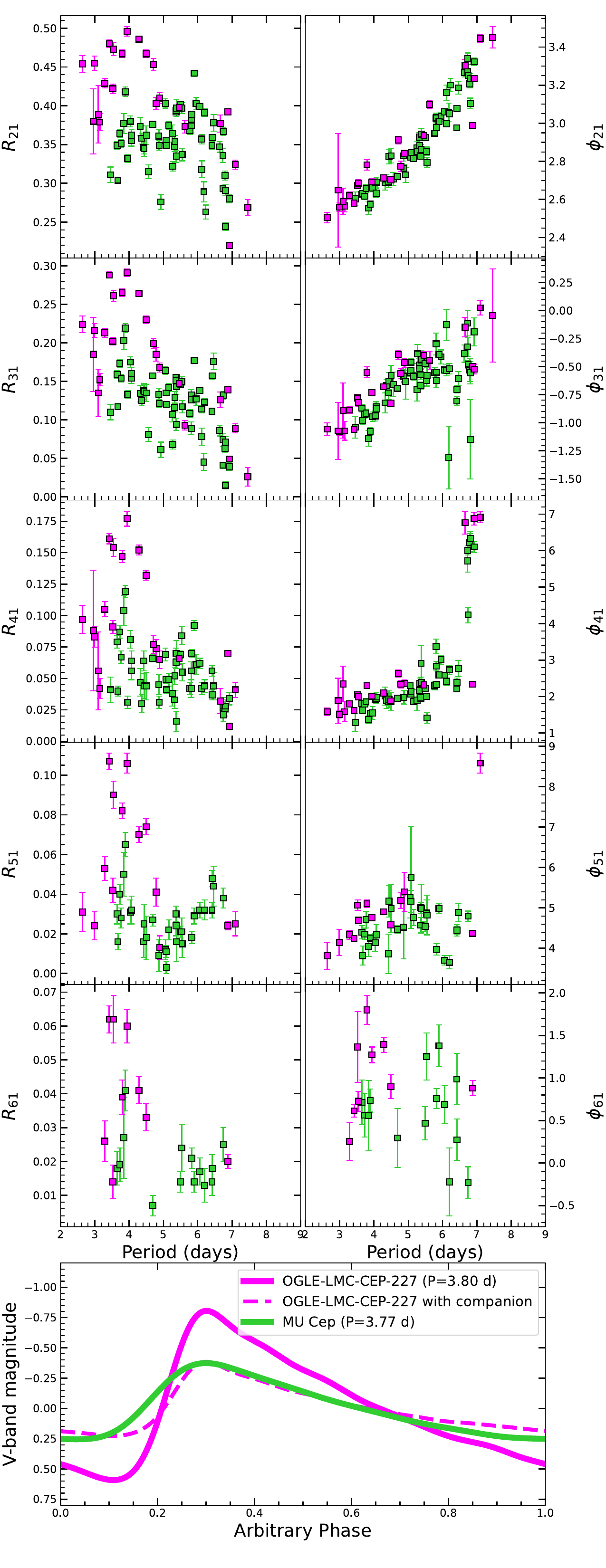}
        \caption{}
        \label{fig:fourier_all_LC}
    \end{subfigure}

    \caption{Fourier amplitude ratios and phase parameters ($R_{21}$ to $R_{61}$ and $\phi_{21}$ to $\phi_{61}$) of Cepheid RV curves (a) and LCs (b) as a function of the period. Models for MU Cep and OGLE-LMC-CEP-227 are shown to illustrate the difference between solar and subsolar metallicity around $P=4\,$day. The solar and sub-solar metallicity variables are plotted with yellow and red error bars (resp. magenta and green) for the RV curves (resp. LCs). Fourier parameters are derived with decomposition analysis presented in Sect.~\ref{sect:fourier}.}
\end{figure*}

The qualitative trends of Fourier parameters are well documented in the literature for both $V$-band LCs \citep{Soszynski2008,Soszynski2010,Bhardwaj2015,Hocde2023} and RV curves \citep{kovacs90,Anderson2024,Hocde2024RV}. We observe first that the Fourier phases exhibit tight progressions with pulsation period for RV curves, while analogous progressions for the LCs are more scattered. We also confirm that there is no break in $\phi_{41}$(RV) around 7$\,$days, contrary to what is observed for the LCs \citep{SimonMoffett1985}, compare $\phi_{41}$ from Figs.~\ref{fig:fourier_all_RV} and  \ref{fig:fourier_all_LC}. The difference between $\phi_{41}$ of the RV curves and LCs was first noted by \cite{kovacs90} who attributed it to unknown photospheric phenomena. 

\begin{figure}[htbp]
    \centering
    \begin{subfigure}[b]{0.24\textwidth}
        \includegraphics[width=\linewidth]{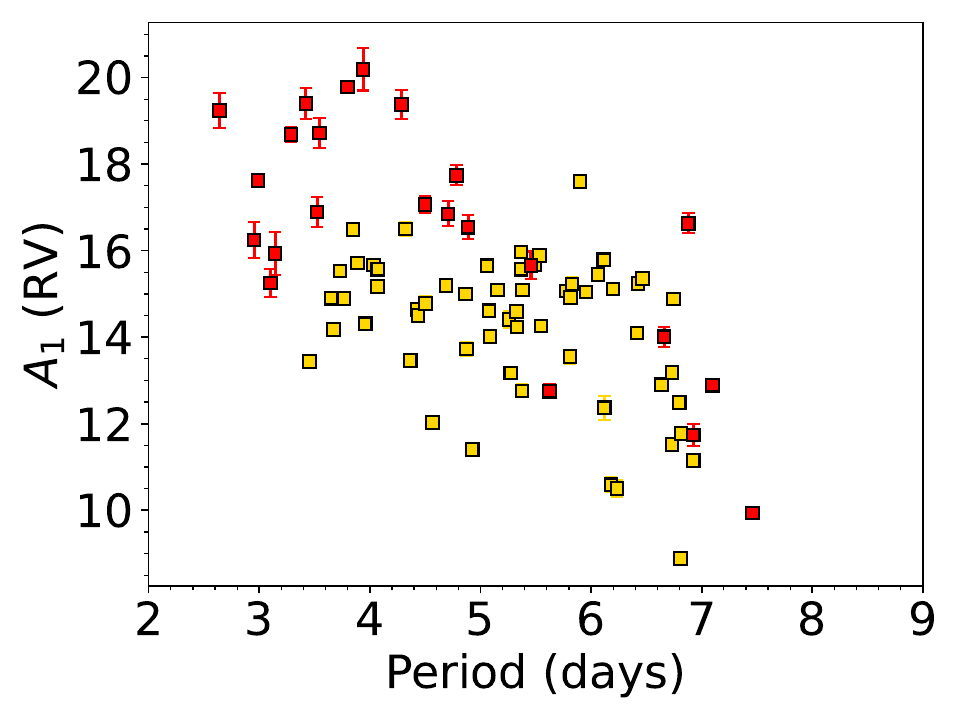}
        \caption{}
        \label{fig:A1_vs_Period_RV}
    \end{subfigure}
    \begin{subfigure}[b]{0.24\textwidth}
        \includegraphics[width=\linewidth]{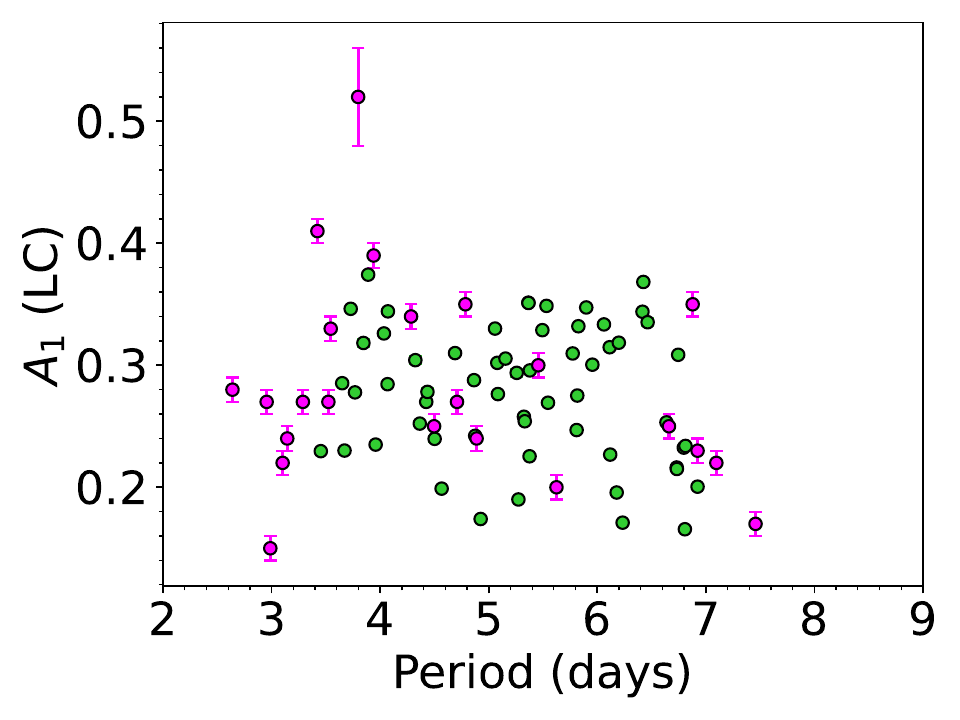}
        \caption{}
        \label{fig:A1_vs_Period_LC}
    \end{subfigure}
    \caption{Amplitude of the first harmonic $A_1$ for RV curves (left panel) and for LCs (right panel). Cepheids of solar metallicity are plotted with yellow and green symbols and those of sub-solar metallicity with red and magenta symbols.}
    \label{fig:A1}
\end{figure}

Contrary to the Fourier phases, amplitude ratios display larger scatter without noticeable trend except for $R_{21}$. In the latter case, RV curves become increasingly asymmetric with the pulsation period, whereas the LCs tend toward a more symmetric profile. As an example, for the Cepheid U~Vul ($P$=7.99\,days) we derived $R_{21}$(LC)=0.228, while $R_{21}$(RV)=0.498, which is more than twice as large. Interestingly, this apparent opposite behaviour of LCs and RV curves is already visible from pulsation models, see Figure 6 in  \cite{Moskalik1992}, and more recently, \cite{Paxton2019}. 
These effects are caused by difference in behavior of the second harmonic $A_2$, since the amplitudes $A_1$(RV) and $A_1$(LC) are scattered in Fig.~\ref{fig:A1}, showing no particular trend with the pulsation period. This is clearly visible for the calibrating sample of the Galactic Cepheids (yellow and green symbols).

Several studies have examined empirically the impact of metallicity on LC shapes of Cepheids \citep{Antonello2000,Klagyivik2007,Szabados2012,Majaess2013,Klagyivik2013,Bhardwaj2015,Hocde2023}. In the case of the short-period Cepheids, the most significant effects are observed in $R_{21}$ and $R_{31}$, which are higher for metal-poor Cepheids, with tight empirical correlations \citep{Klagyivik2013,Hocde2023}. For RV curves, \citet{Pont2001} found that metal-poor Cepheids with periods below 5\,days show larger values of $A_1$ and $R_{21}$, while $\phi_{21}$ remains unaffected. We confirm these trends using a sample of metal-poor Cepheids twice as large, after the inclusion of the RV curves of the LMC and SMC stars (see red error bars in Figs.~\ref{fig:fourier_all_RV} and~\ref{fig:A1}). We also observe a prominent peak in the amplitude ratios at around 4\,days for both LCs and RVs of metal-poor Cepheids. Moreover, we show for the first time that this effect is persistent for amplitude ratios up to the 7th order, while the Fourier phases remain unaffected. 

These qualitative descriptions highlight the influence of both pulsation period and metallicity on the shape of the RV curves in short-period Cepheids. The clear and regular progression of the RV Fourier phases makes them easier to model as a function of pulsation period. In contrast, the significant scatter and the apparent discrepancies between $R_{21}$ of LCs and RV curves motivate us to investigate potential correlations between them, with particular attention to the role of metallicity.

\section{Reconstruction of RV curves from the LC shapes}\label{sect:templates}
\subsection{Empirical relations}
In order to reconstruct the RV curves of Cepheids, we need to infer the Fourier parameters of these RV curves. This is straightforward in the case of the Fourier phase differences, since these parameters follow a well-defined relation with the pulsation period. For the amplitudes and amplitude ratios of the RV curves, we explored possible correlations with the Fourier parameters of the LCs. Specifically, we first visually inspected various combinations of $R_{k1}$(RV) as a function of $R_{k1}$(LC) and the pulsation period, and $A_1$(RV) as a function of $A_1$(LC) or $R_{k1}$(LC). As a result, we find several empirical relations up to order seven for amplitude ratios and phase differences as presented in Fig.~\ref{fig:correlations}, and for the amplitude $A_1$ in Fig.~\ref{fig:fit_A1}. 
In many of these empirical relations we observe a
break at a period of 7\,days. For this reason we
restrict our subsequent analysis to Cepheids with $P <7\,$day and omit the ones with longer periods (plotted with grey symbols in Figs.~\ref{fig:correlations} and \ref{fig:fit_A1}).

 In order to model the different correlations, we chose a simple approach of fitting either linear relations of the form
\begin{equation}
   Y= aX+b
\end{equation}
or a power-law
\begin{equation}
   Y= aX^b+c
\end{equation}
For all relations between different Fourier parameters, we applied Orthogonal Distance Regression (ODR) to properly account for uncertainties in both the dependent and independent variables. This method minimizes the orthogonal distances between data points and the fitted model, thereby incorporating errors in both variables. We utilized the ODR implementation provided by the SciPy library\footnote{\url{https://docs.scipy.org/doc/scipy/reference/odr.html}}. For the empirical relations depending on the pulsation period, we employed a standard nonlinear least-squares fitting, using the \texttt{curve\_fit} function\footnote{\url{https://docs.scipy.org/doc/scipy/reference/generated/scipy.optimize.curve_fit.html}} from Python, considering only the uncertainties in the dependent variable. 

We display the fitted relations with blue lines in  Figs.~\ref{fig:correlations} and \ref{fig:fit_A1} 
and we provide the coefficients with statistical errors, reduced $\chi^2_r$ and RMS of the fits in Table~\ref{tab:fourier_relations_full}.

As expected, phase differences correlate well with the pulsation period as we can see from Figs~\ref{fig:phi21} to \ref{fig:phi71}. For the amplitude ratios we discovered a progression of the ratio of amplitude ratios $R_{21}$(RV)/$R_{21}$(LC) with the pulsation period, and similarly for $R_{31}$(RV)/$R_{31}$(LC) (see Figs.~\ref{fig:R21} and \ref{fig:R31}). This result contrasts with the apparent large scatter of $R_{21}$ and $R_{31}$ described in the previous section, and shows that the shape of both the LCs and RV curves follows empirical correlations with the pulsation period. Furthermore, we find that all higher-order $R_{k1}$(RV) parameters correlate well with $R_{21}$(LC) up to order seven (see Figs.~\ref{fig:R41} to \ref{fig:R71}). Last, $A_1$(RV) correlates with $R_{21}$(LC), $R_{31}$(LC) and a combination of both (see Figs.~\ref{fig:A1_R21} to \ref{fig:A1_R21R31}).

\subsection{Metallicity dependence}
\begin{figure*}[!htbp]
\centering
\begin{subfigure}[b]{.35\textwidth}
  \includegraphics[width=\linewidth]{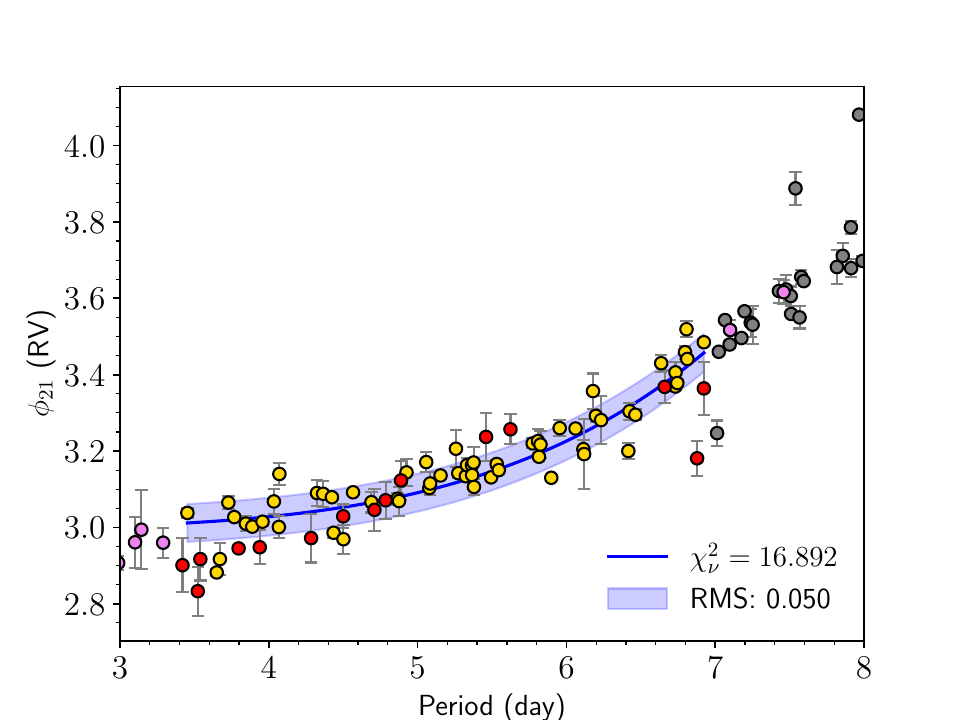}
  \caption{}
  \label{fig:phi21}
\end{subfigure}%
\begin{subfigure}[b]{.35\textwidth}
  \includegraphics[width=\linewidth]{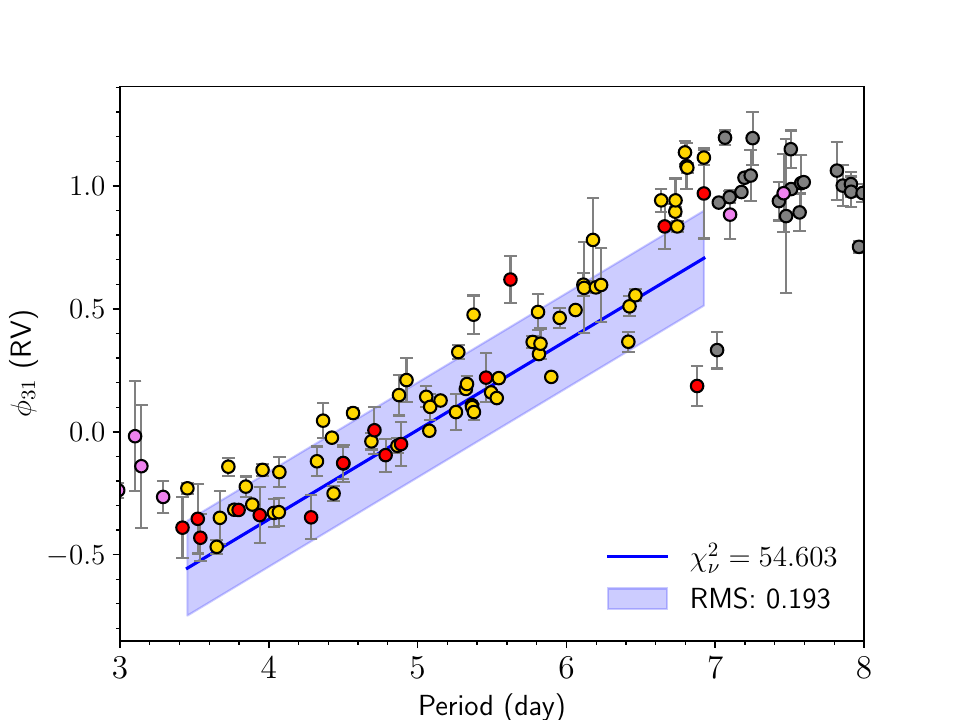}
  \caption{}
  \label{fig:phi31}
\end{subfigure}%
\begin{subfigure}[b]{.35\textwidth}
  \includegraphics[width=\linewidth]{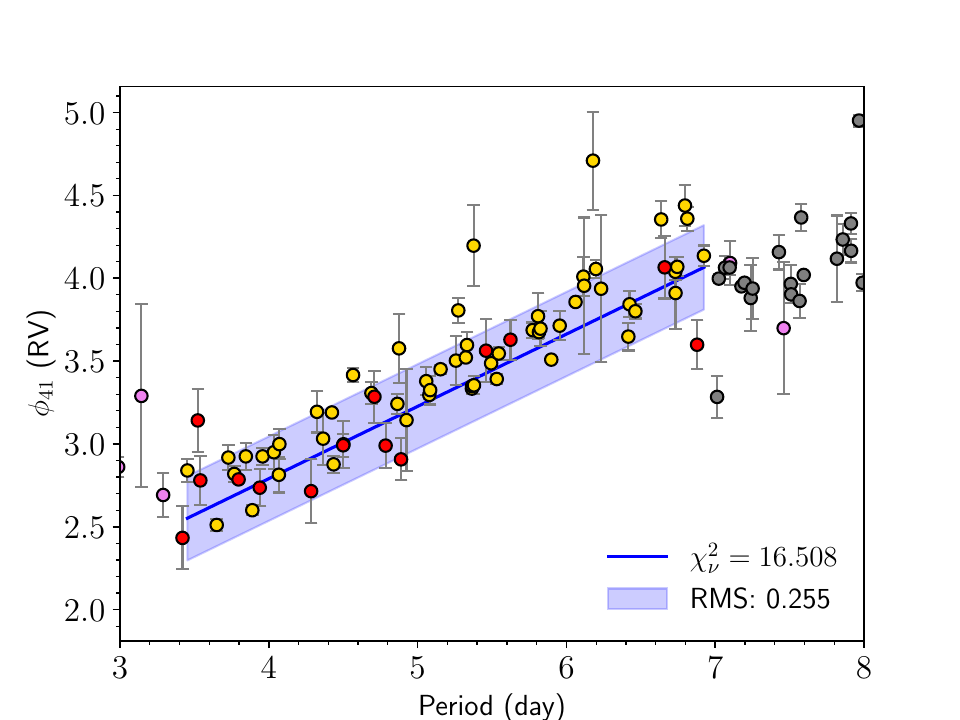}
  \caption{}
  \label{fig:phi41}
\end{subfigure}\vskip\baselineskip
\begin{subfigure}[b]{.35\textwidth}
  \includegraphics[width=\linewidth]{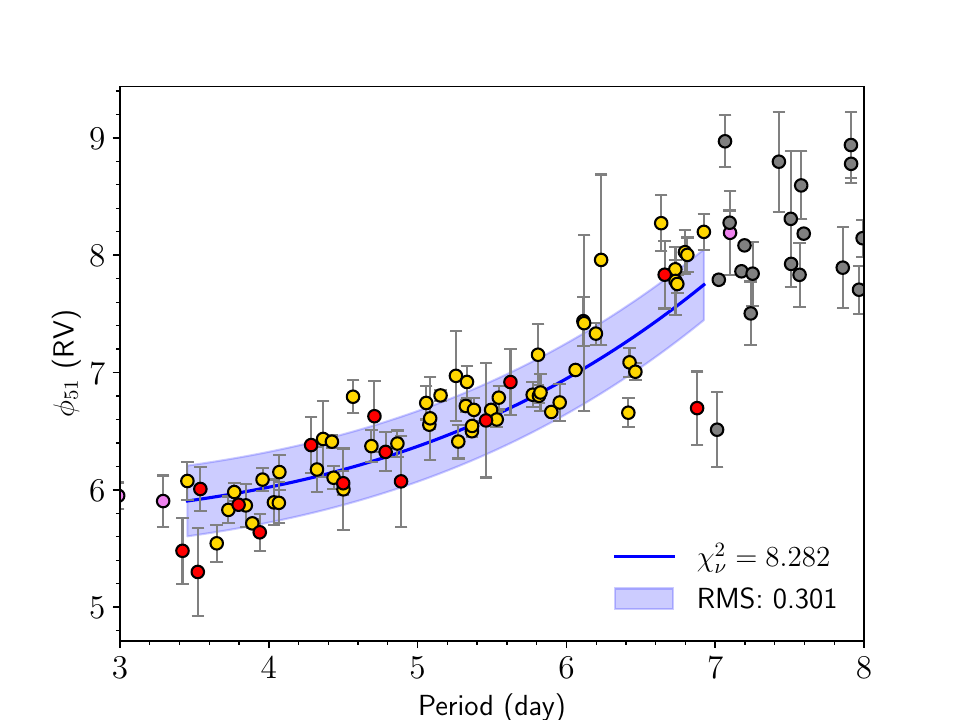}
  \caption{}
  \label{fig:phi51}
\end{subfigure}%
\begin{subfigure}[b]{.35\textwidth}
  \includegraphics[width=\linewidth]{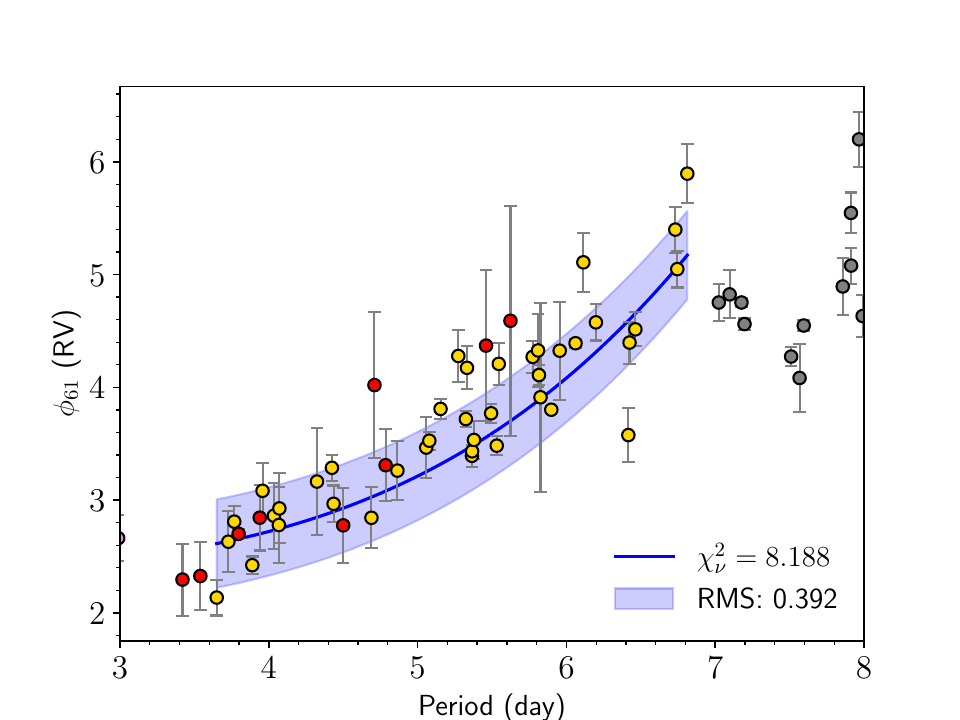}
  \caption{}
  \label{fig:phi61}
\end{subfigure}%
\begin{subfigure}[b]{.35\textwidth}
  \includegraphics[width=\linewidth]{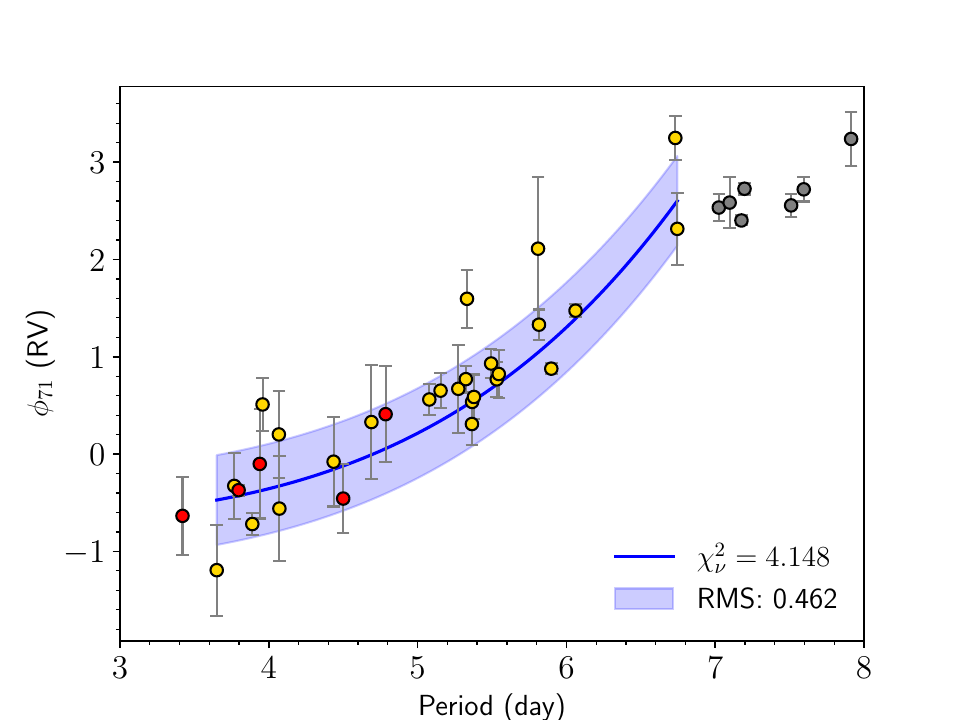}
  \caption{}
  \label{fig:phi71}
\end{subfigure}\vskip\baselineskip
\begin{subfigure}[b]{.35\textwidth}
  \includegraphics[width=\linewidth]{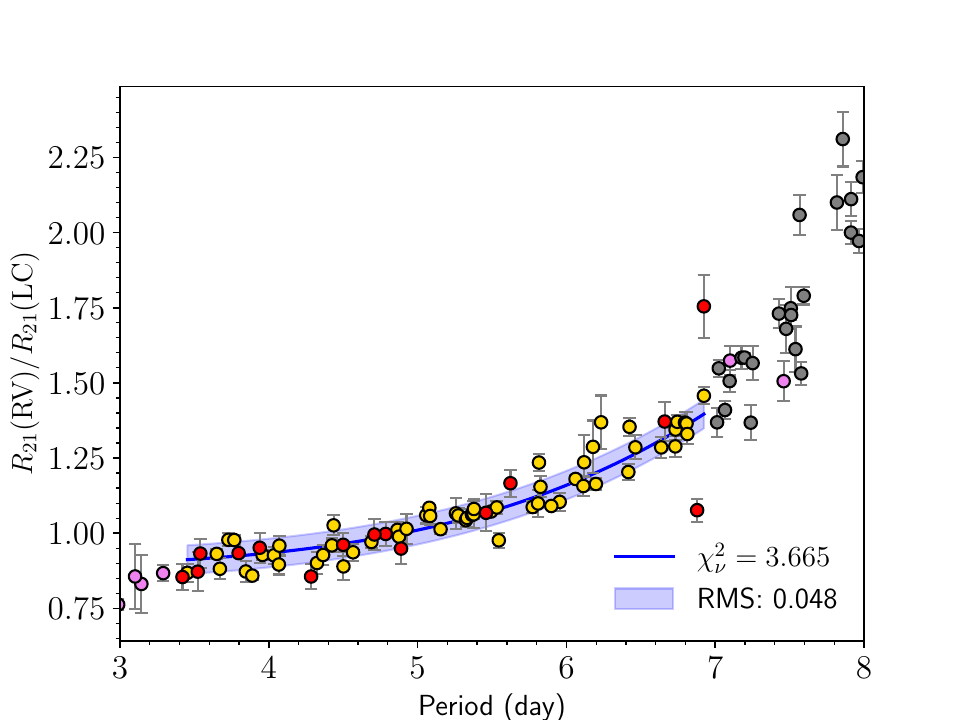}
  \caption{}
  \label{fig:R21}
\end{subfigure}%
\begin{subfigure}[b]{.35\textwidth}
  \includegraphics[width=\linewidth]{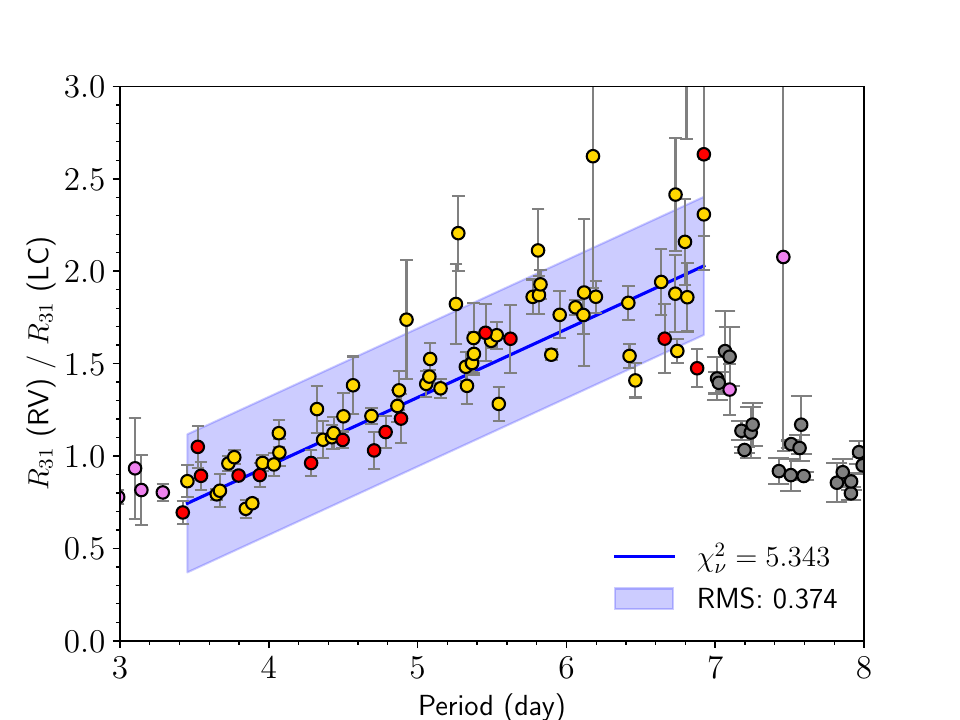}
  \caption{}
  \label{fig:R31}
\end{subfigure}%
\begin{subfigure}[b]{.35\textwidth}
  \includegraphics[width=\linewidth]{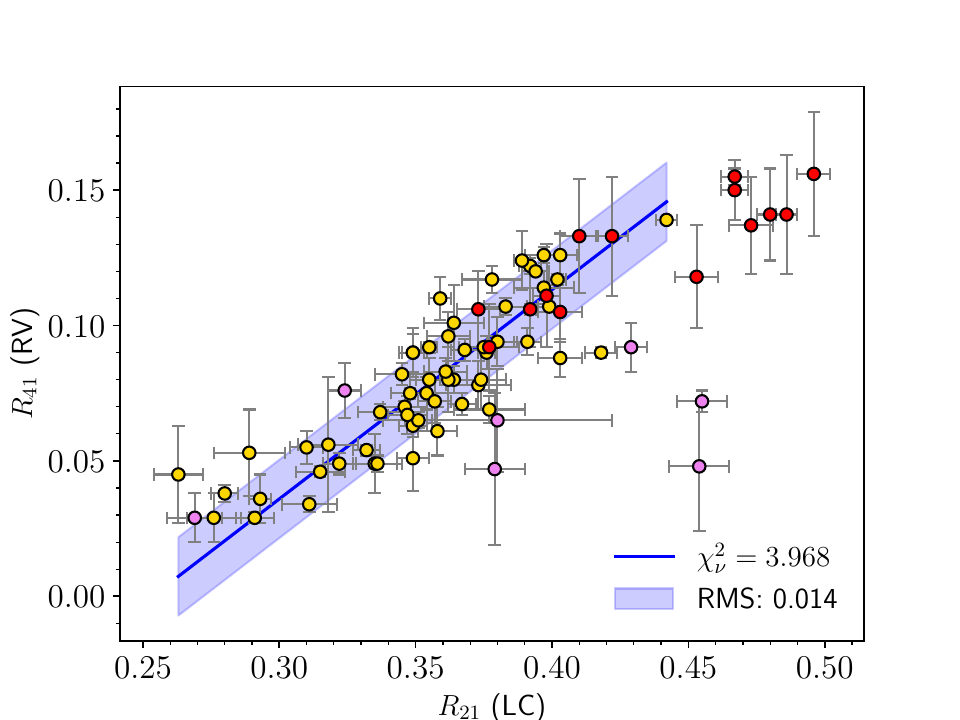}
  \caption{}
  \label{fig:R41}
\end{subfigure}\vskip\baselineskip
\begin{subfigure}[b]{.35\textwidth}
  \includegraphics[width=\linewidth]{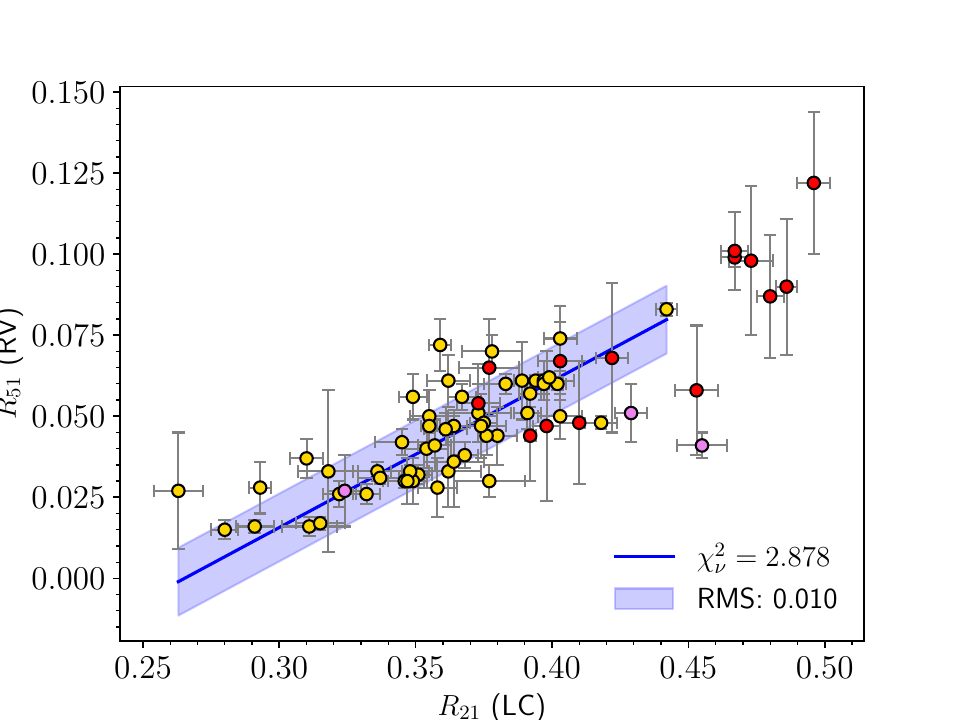}
  \caption{}
  \label{fig:R51}
\end{subfigure}%
\begin{subfigure}[b]{.35\textwidth}
  \includegraphics[width=\linewidth]{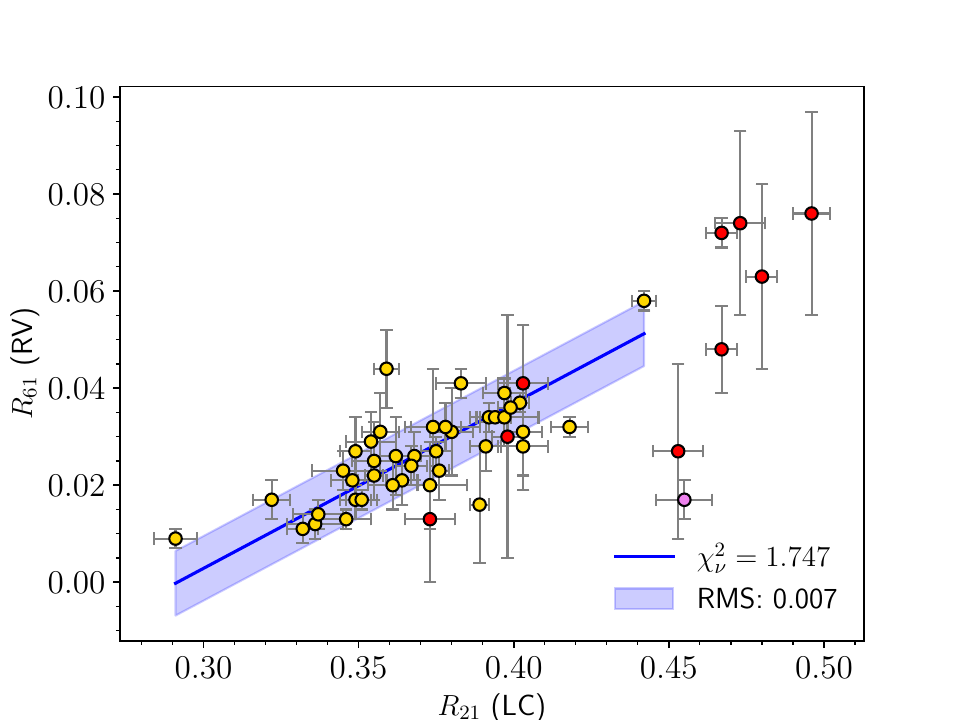}
  \caption{}
  \label{fig:R61}
\end{subfigure}%
\begin{subfigure}[b]{.35\textwidth}
  \includegraphics[width=\linewidth]{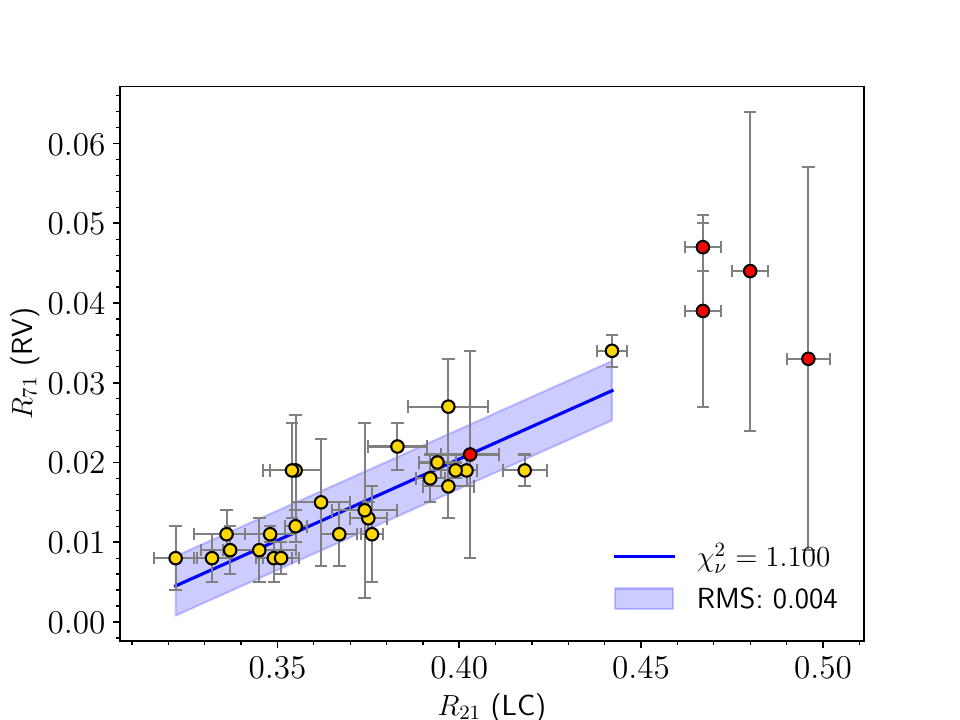}
  \caption{}
  \label{fig:R71}
\end{subfigure}
\caption{Fourier parameters of RV curves as a function of pulsation period and $V$-band LC shape: 
(a)~$\phi_{21}$, (b)~$\phi_{31}$, (c)~$\phi_{41}$, 
(d)~$\phi_{51}$, (e)~$\phi_{61}$, (f)~$\phi_{71}$, 
(g)~$R_{21}$, (h)~$R_{31}$, (i)~$R_{41}$, 
(j)~$R_{51}$, (k)~$R_{61}$, (l)~$R_{71}$. 
Yellow circles (calibrating sample) are used to produce the fits between 3.45 and 7\,days (blue lines, see Sect.~\ref{sect:templates}); red circles show metal-poor Cepheids; grey points above $P > 7$\,days are ignored in the fit; violet points are metal-poor Cepheid outside the pulsation period range 3.4 to 7 days.}
\label{fig:correlations}
\end{figure*}

\begin{figure*}[!htbp]
    \centering
    \begin{subfigure}[b]{0.33\textwidth}
        \centering
        \includegraphics[width=\linewidth]{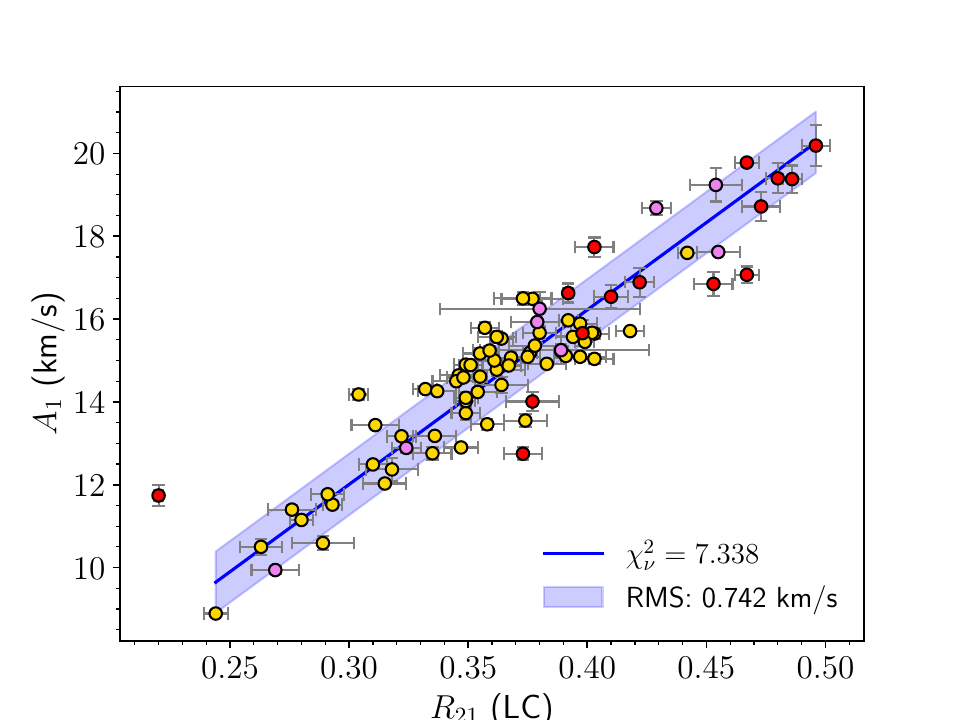}
        \caption{}
        \label{fig:A1_R21}
    \end{subfigure}
    \begin{subfigure}[b]{0.33\textwidth}
        \centering
        \includegraphics[width=\linewidth]{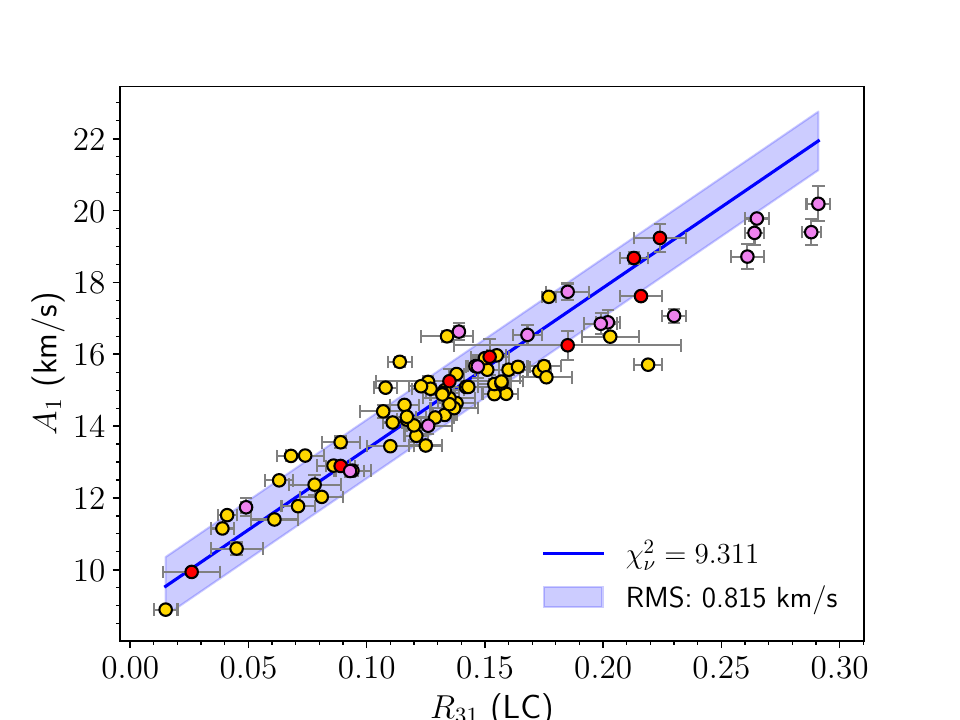}
        \caption{}
        \label{fig:A1_R31}
    \end{subfigure}
        \begin{subfigure}[b]{0.33\textwidth}
        \centering
        \includegraphics[width=\linewidth]{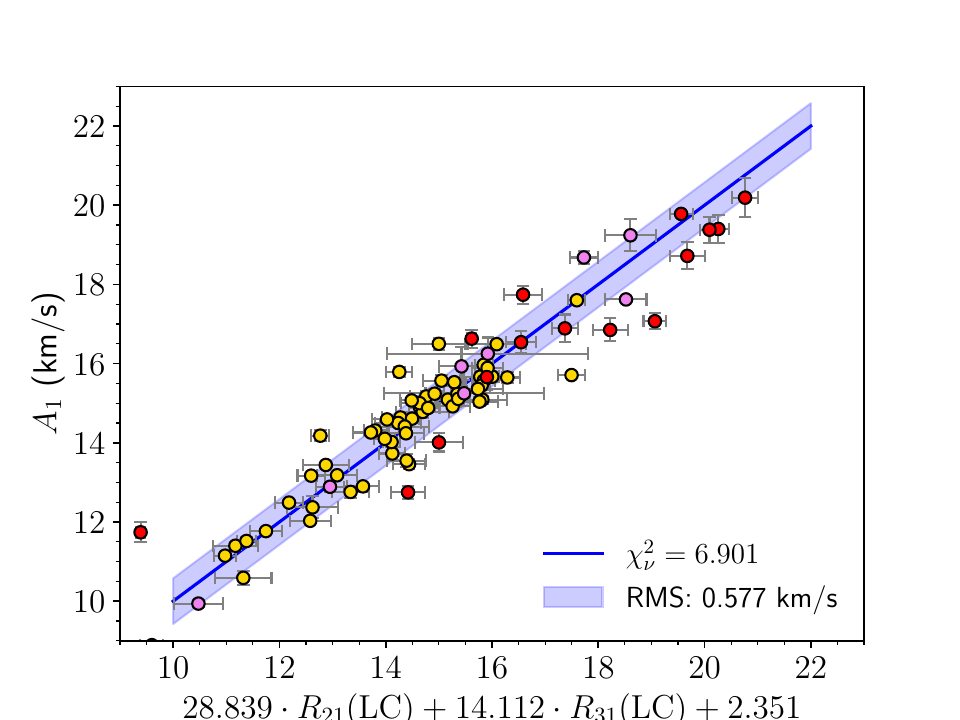}
        \caption{}
        \label{fig:A1_R21R31}
    \end{subfigure}
        \caption{Fourier parameter $A_1$ of the RV curves as a function of (a) $R_{21}$(LC), (b) $R_{31}$(LC) and (c) a combination of $R_{21}$(LC) and $R_{31}$(LC). The ODR fit (blue line) is applied to the calibrating sample (yellow circles) and prolongated to show consistency with metal-poor stars (red circles). }\label{fig:fit_A1}
\end{figure*}

For all these empirical relations, we overplot Fourier parameters of metal-poor Cepheids in red in Figs.~\ref{fig:correlations} and \ref{fig:fit_A1}. These Fourier parameters are derived from RV curves of much lower quality and thus must be interpreted with caution. Despite the lower accuracy, we found that metal-poor Cepheids follow closely the empirical relations $R_{21}$(RV)/$R_{21}$(LC)  and  $R_{31}$(RV)/$R_{31}$(LC) between 3.5 and 7\,days with the exception of two outliers (OGLE-LMC-1327 and 1249).

We also find that the $A_1$(RV) coefficient of metal-poor Cepheids follows very closely the linear correlation with $R_{21}$(LC), $R_{31}$(LC) or a combination of both, defined by our calibrating sample (see Fig.~\ref{fig:fit_A1}). Despite larger error bars, we also see that the higher-order amplitude ratios of metal-poor Cepheids follow the trend of our calibrating sample. 
Although the lower precision of the Fourier parameters of metal-poor RV curves, combined with the limited size of the sample, prevents us from making a firm conclusion, our results indicate that the empirical relations presented in this paper do not depend strongly on the metallicity.

\begin{table*}[ht]
\centering
\caption{Empirical relations for the Fourier phases $\phi_{k1}(\mathrm{RV})$ and amplitude ratios $R_{k1}$(RV) as functions of the pulsation period $P$ or LC Fourier parameters $R_{21}$ and $R_{31}$.}
\begin{tabular}{l @{$\quad\quad$} l @{$\ \ $} |l @{$\quad$} |l @{$\quad$}}
\hline
\multicolumn{2}{c}{Fourier phase relations} & RMS & $\chi^2_r$\\
\hline
$\phi_{21}(\mathrm{RV})$ &= $(0.00004 \pm 0.00001)\cdot \mathrm{P}^{(4.853 \pm 0.139)} + (2.996 \pm 0.005)$ & 0.050 &16.89\\
$\phi_{31}(\mathrm{RV})$ &= $ (0.363 \pm 0.003) \cdot \mathrm{{P}}+(-1.809 \pm 0.017)$ & 0.193&54.60\\
$\phi_{41}(\mathrm{RV})$ &= $(0.436 \pm 0.007) \cdot \mathrm{P} + (1.046 \pm 0.038)$ & 0.255&16.51\\
$\phi_{51}(\mathrm{RV})$ &= $(0.002 \pm 0.001) \cdot \mathrm{{P}}^{{3.534 \pm 0.259}} + (5.733 \pm 0.066)$ & 0.301&8.28\\
$\phi_{61}(\mathrm{RV})$ &= $(0.0017 \pm 0.0012) \cdot \mathrm{P}^{(3.863 \pm 0.361)} + (2.363 \pm 0.115)$ & 0.392 &8.19\\
$\phi_{71}(\mathrm{RV})$ &= $(0.0010 \pm 0.0011) \cdot \mathrm{P}^{(4.242 \pm 0.549)} + (-0.718 \pm 0.186)$ & 0.462&4.15\\
\hline
\multicolumn{2}{c}{Amplitude relations} & RMS & $\chi^2_r$\\
\hline
$A_1(\mathrm{RV})$ &= $(42.151 \pm 2.307) \cdot R_{21}(\mathrm{{LC}})+(-0.638 \pm 0.834)$ & 0.742\,km/s&7.34\\
$A_1(\mathrm{RV})$ &= $(44.936 \pm 2.813) \cdot R_{31}(\mathrm{{LC}})+(8.865 \pm 0.376)$ & 0.815\,km/s &9.31\\
$A_1(\mathrm{RV})$ &= $(28.839 \pm 3.078) \cdot R_{21}(\mathrm{LC}) + (14.112 \pm 3.051) \cdot R_{31}(\mathrm{LC}) + (2.351 \pm 0.821)$ & 0.577\,km/s&6.90\\
$R_{21}(\mathrm{RV})/{R_{21}(\mathrm{LC})}$ &= $(0.00011 \pm 0.00006) \cdot \mathrm{{P}}^{{4.373 \pm 0.281}} + (0.888 \pm 0.011)$ & 0.048&3.66\\
$R_{31}(\mathrm{RV})/{R_{31}(\mathrm{LC})}$ &= $(0.370 \pm 0.009) \cdot \mathrm{{P}} + (-0.534 \pm 0.043)$ & 0.374&5.34\\
$R_{41}(\mathrm{RV})$ &= $(0.773 \pm 0.045) \cdot R_{21}(\mathrm{LC}) + (-0.196 \pm 0.017)$ & 0.014&3.97\\
$R_{51}(\mathrm{RV})$ &= $(0.452 \pm 0.028) \cdot R_{21}(\mathrm{LC}) + (-0.120 \pm 0.010)$ & 0.010&2.88\\
$R_{61}(\mathrm{RV})$ &= $(0.341 \pm 0.024) \cdot R_{21}(\mathrm{LC}) + (-0.099 \pm 0.009)$ & 0.007&1.75\\
$R_{71}(\mathrm{RV})$ &= $(0.204 \pm 0.019) \cdot R_{21}(\mathrm{LC}) + (-0.061 \pm 0.007)$ & 0.004&1.10\\
\hline
\end{tabular}
\label{tab:fourier_relations_full}
\end{table*}

\subsection{Reconstruction method}
With the empirical relations established in the previous section, it is then possible to reconstruct the RV curve of a given Cepheid using the Fourier parameters of the $V$-band LC and the pulsation period. Fourier phase differences $\phi_{k1}$(RV) up to order 7 can be determined knowing the pulsation period only. Fixing arbitrarily $\phi_1=0$, we are then able to unfold all phases $\phi_{k>1}$(RV) using Eq.~4 and fitted relations in Table~\ref{tab:fourier_relations_full}. On the other hand,  the amplitude ratios $R_{21}$(LC) and $R_{31}$(LC) of the $V$-band LC allow the determination of the amplitude ratios of the RV curves $R_{k1}$(RV) up to order 7 as well as the amplitude of the first harmonic $A_1$(RV). We adopt the $A_1$ relation as a function of the combination of  $R_{21}$(LC) and $R_{31}$(LC), as it provides the smaller RMS with 0.577\,km/s. Then, we deduce the amplitudes of all the remaining harmonic since $A_k=A_1R_{k1}$.
As a result, we can sum up all the harmonics to reconstruct the RV curves $V^\mathrm{rec}_r(t)$ following
\begin{equation}
V^\mathrm{rec}_r(t) = \sum_{k=1}^n A_k \mathrm{sin}[k \omega t +\phi_k]
,\end{equation}

\begin{figure*}[]
\begin{subfigure}[b]{.33\textwidth}
  \includegraphics[width=\linewidth]{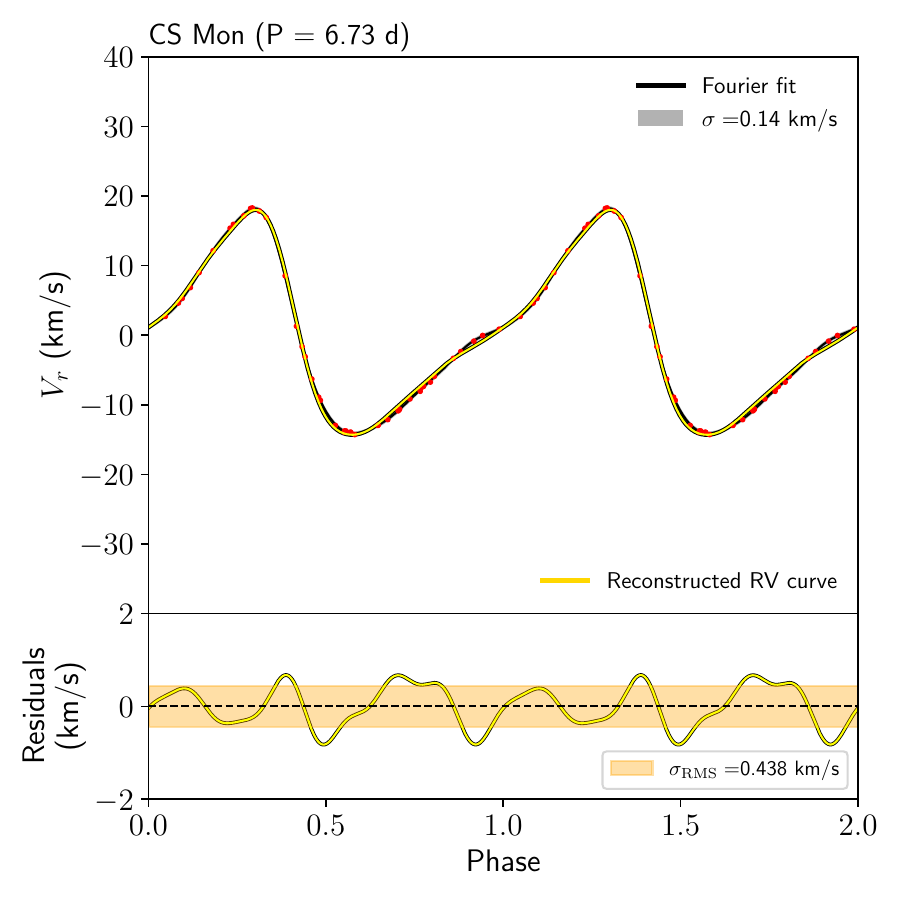}
  \caption{}
  \label{fig:cs_mon}
\end{subfigure} 
\begin{subfigure}[b]{.33\textwidth}
  \includegraphics[width=\linewidth]{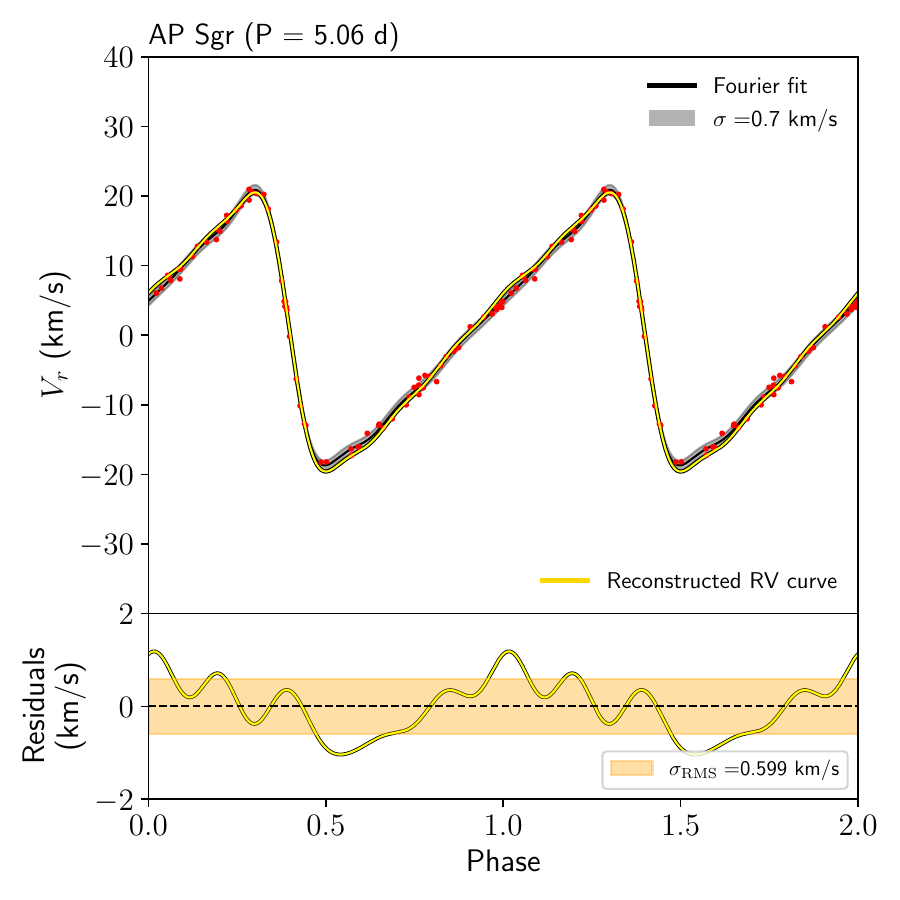}
  \caption{}
  \label{fig:ap_sgr}
\end{subfigure}
\begin{subfigure}[b]{.33\textwidth}
  \includegraphics[width=\linewidth]{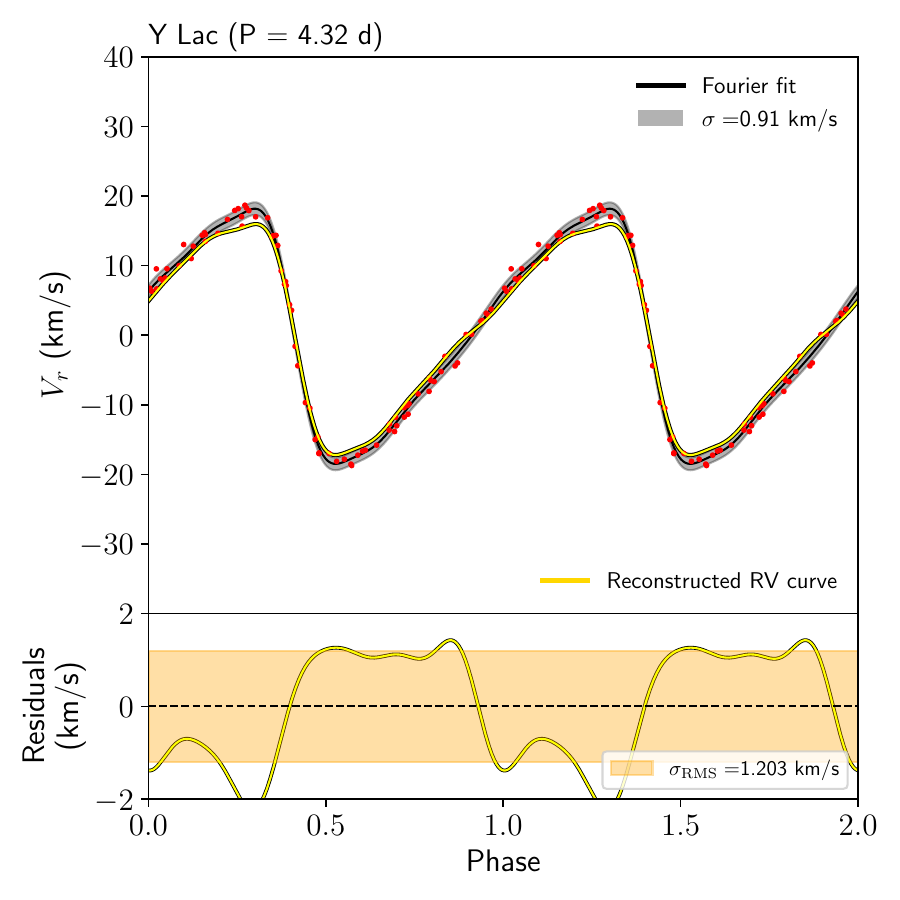}
  \caption{}
  \label{fig:y_lac}
\end{subfigure} 
\caption{Examples of reconstructed RV curves of excellent, medium and lower accuracy in the case of (a) CS Mon, (b) AP Sgr and (c) Y Lac, respectively. The yellow curve shows the reconstructed RV curve using empirical relations defined in Table \ref{tab:fourier_relations_full}, the black curve is the Fourier fit to the RV observations (red points, see Table~\ref{tab:data_ref}). The accuracy of the reconstructed RV curve is defined by the standard  deviation of the residuals along the pulsation cycle.}
 \label{fig:sample_RV}
\end{figure*}

\begin{figure*}[]
\begin{subfigure}[b]{.33\textwidth}
  \includegraphics[width=\linewidth]{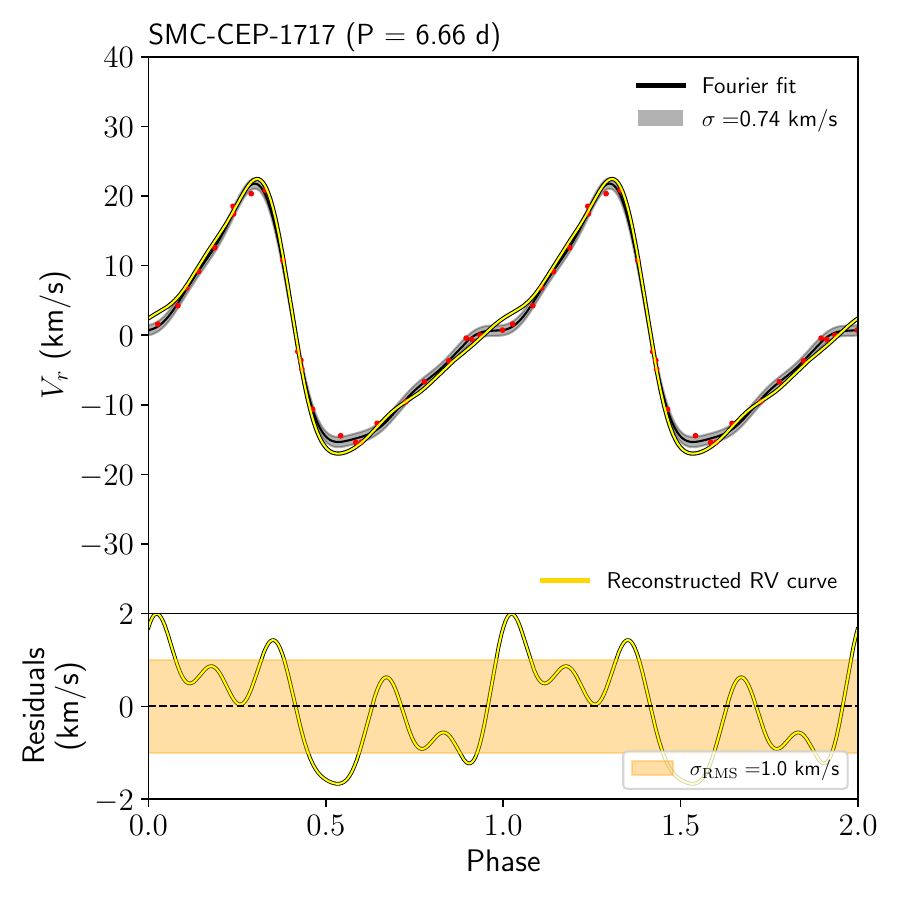}
  \caption{}
  \label{fig:smc-1717}
\end{subfigure} 
\begin{subfigure}[b]{.33\textwidth}
  \includegraphics[width=\linewidth]{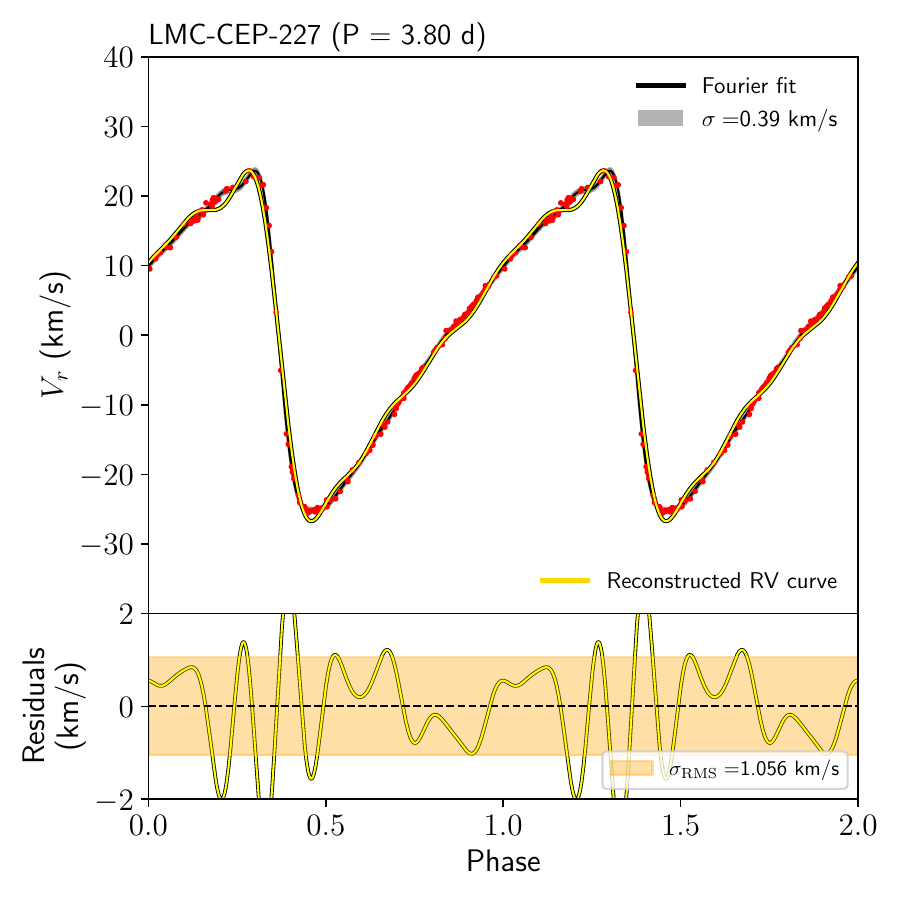}
  \caption{}
  \label{fig:lmc-227}
\end{subfigure}
\begin{subfigure}[b]{.33\textwidth}
  \includegraphics[width=\linewidth]{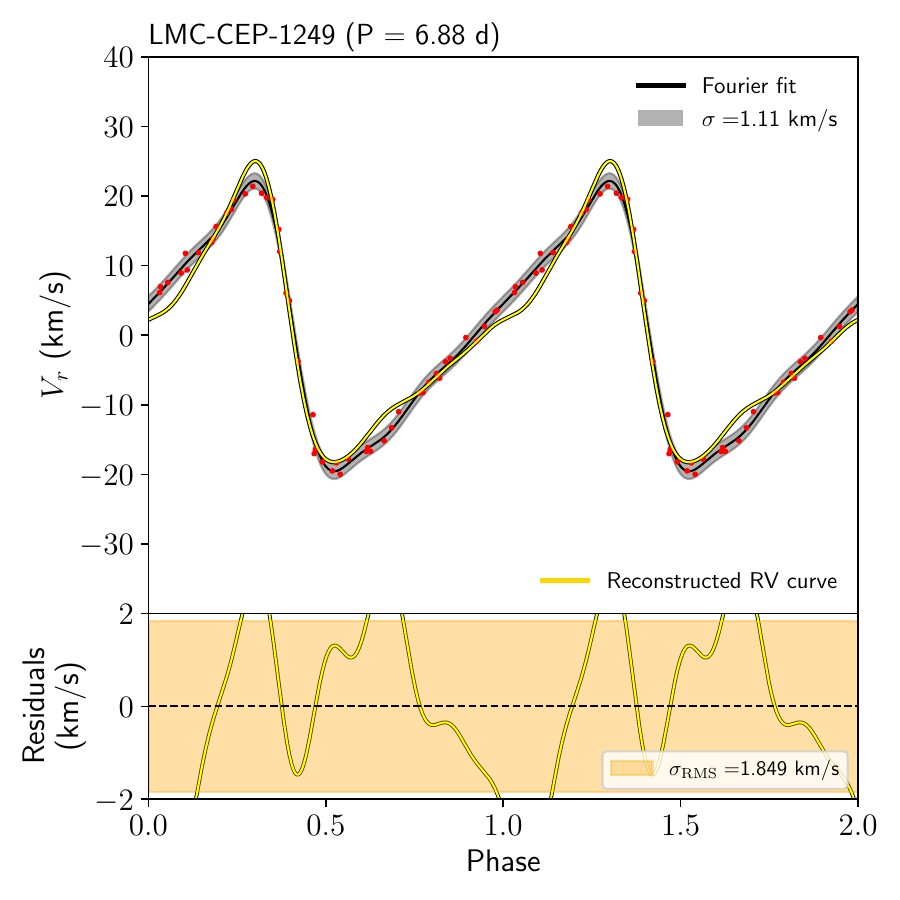}
  \caption{}
  \label{fig:lmc-1249}
\end{subfigure} 
\caption{Examples of reconstructed metal-poor RV curves of good, medium and lower accuracy for the Cepheids (a) OGLE-SMC-1717, (b) OGLE-LMC-227 and (c) OGLE-LMC-1249. The black curve is the Fourier fit to the RV observations (red points, see Table~\ref{tab:data_ref_metal_poor}).}
 \label{fig:sample_RV_metal-poor}
\end{figure*}

Analyzing the accuracy of reconstruction for different orders of the Fourier fit, we determined $n=6$ as the optimal order for our calibrating sample. This analysis is presented in Sect.~\ref{sect:order}.
In Fig.~\ref{fig:sample_RV} we provide examples of reconstructed RV curves for several Cepheids of the calibrating sample, and in Fig.~\ref{fig:sample_RV_metal-poor}, for lower quality metal-poor Cepheids. The entire set of reconstructed RV curves is available online on zenodo. The Cepheid CR~Cep is rejected from this reconstruction because the LC is modeled with 2 orders only, thus it is not possible to use the $R_{31}(RV)/R_{31}(LC)$ relation.

\section{Accuracy of the reconstructed RV curves}\label{sect:uncertainties}

In this section, we first estimate the internal accuracy of the reconstructed shape of the RV curves of the calibrating sample, which serve as indicators of the fit quality. We then estimate realistic external accuracy with a Monte Carlo resampling, which enables us to assess how well our method generalizes to stars outside the calibrating sample.

\subsection{Reconstruction residuals}
Our objective is to assess the precision of the reconstructed RV curves with respect to the real RV curves. To this end, our strategy is to compare the RV reconstruction for each star with its RV Fourier fit, defined by the true spectroscopic measurements. This is possible and reliable because the Fourier fits for Cepheids of the calibration sample are of excellent quality and without any significant instabilities \citep{Hocde2024RV}. However, the RV curves of metal-poor Cepheids are not suitable for this comparison because of lower quality fits. Nevertheless, we decided to perform this calculations for metal-poor Cepheids as well, but these results must be treated with caution.

Hence, we compared for each star the reconstructed RV curve, $V^{rec}_r$, with the Fourier fit of the true RV curve, $V^{fit}_r$. This was done by deriving the root mean square of the residuals

\begin{equation}\label{eq:sigma_vr}
    \sigma_{rms}^2(V_r)=\frac{1}{N}\sum_{i=1}^{N}\left(V_r^\mathrm{rec}(\phi_i) -V_r^\mathrm{fit}(\phi_i) \right)^2,
\end{equation}
which is a measure of internal accuracy of the reconstructed RV
   curve.
Since $V_r^\mathrm{rec}$ and $V_r^\mathrm{fit}$ have arbitrary phasing, we chose the phase of $V_r^\mathrm{rec}$ in such way as to minimize the residuals $\sigma_{rms}(V_r)$. The two curves are sampled at $N=100$ regularly spaced phases along the pulsation cycle. The resulting reconstruction residuals for individual stars are plotted vs. the pulsation period in Fig.~\ref{fig:residual_period}. Overall, we find a median $\sigma_{rms}=0.53\,$km/s for the calibrating sample (see orange dashed line in Fig.~\ref{fig:residual_period}), which shows the excellent precision of the reconstructed RV curves. From Fig.~\ref{fig:residual_period} we do not find any trends with the pulsation period. For several stars, the accuracy is even better than 0.5\,km/s, as we can see for example in the case of CS~Mon in Fig.~\ref{fig:cs_mon} with $\sigma_{rms}=0.44\,$km/s. This result is particularly remarkable given that only information on the LC and the pulsation period was used to reconstruct the RV curves.

On the other hand, five stars have reconstruction residuals of 1\,km/s or more: AW~Per, FM~Aql, Y~Lac, SS~Sct and CS~Ori. This is most likely caused by a mismatch of the amplitudes, like in the case of Y Lac (see Fig.~\ref{fig:y_lac}), because all these stars except of AW Per are outliers of the $A_1$ relation (see Fig.~\ref{fig:A1_R21R31}). At this stage, we do not have a clear explanation for the origin of the $A_1$(RV) outliers. We note, however, that the derived empirical relations for $A_1$ are highly sensitive to the LC amplitude ratios $R_{21}$ and $R_{31}$. Consequently, small errors in the numerical determination of these parameters, or photometric contamination of the light curve, could bias the estimation of $A_1$. According to our calibration, an offset as small as $\Delta R_{21}(\mathrm{LC}) = 0.035$ can bias $A_1(\mathrm{RV})$ by about 1\,km.s$^{-1}$.

It is well known that blending with a bright companion or crowding can bias both the mean brightness and the LC amplitude \citep{Mochejska2000,KlagyivikSzabados2009}. Several Cepheids in our sample are known binaries \citep{KlagyivikSzabados2009,Shetye2024}. Among five objects with reconstruction residuals above 1km/s, binarity is well established for AW Per \cite{Evans2024} and has been suggested for CS Ori \citep{SzabadosPont1998}. Presence of a photometric companion has also been claimed in case of Y Lac \citep{Evans1990}.

However, we emphasize that blending has a relatively small effect on the LC amplitude ratios, as shown by \citet{Antonello2002}: a strong blending of about 0.45\,mag, equivalent to 50\% of the Cepheid luminosity, is required to increase $R_{21}$ by only 0.02 (see their Figure 2). Although such bright companions can exist, we know from analysis of the PL relation that they are rare \citep{Pilecki2021}. Another possible source of photometric contamination could arise from circumstellar emission during the pulsation cycle in the $V$-band \citep{Hocde2020a,Hocde2020b}.

As expected, the reconstruction of the RV curves for metal-poor Cepheids yields significantly larger residuals than for the calibrating Cepheids, with a median of $\sigma_{rms} = 1.09\,$km/s (see red crosses in Fig.~\ref{fig:residual_period}). We emphasize that in most of our metal-poor Cepheids, the Fourier fits to the true spectroscopic RV curves are of lower quality and as such they are not accurate enough to properly assess the uncertainties of the reconstruction method. Despite this shortcoming, for several metal-poor Cepheids the reconstructed RV curves differ by less than 1\,km/s from the true RV curves (XX Mon, BC Pup, OGLE-SMC-1680 and OGLE-SMC-1717 also presented in Fig.~\ref{fig:smc-1717}).

\subsection{Accuracy of the integrated RV curves}
\begin{figure*}[htbp]
\centering

    \begin{subfigure}[b]{0.45\textwidth}
    \includegraphics[width=\textwidth]{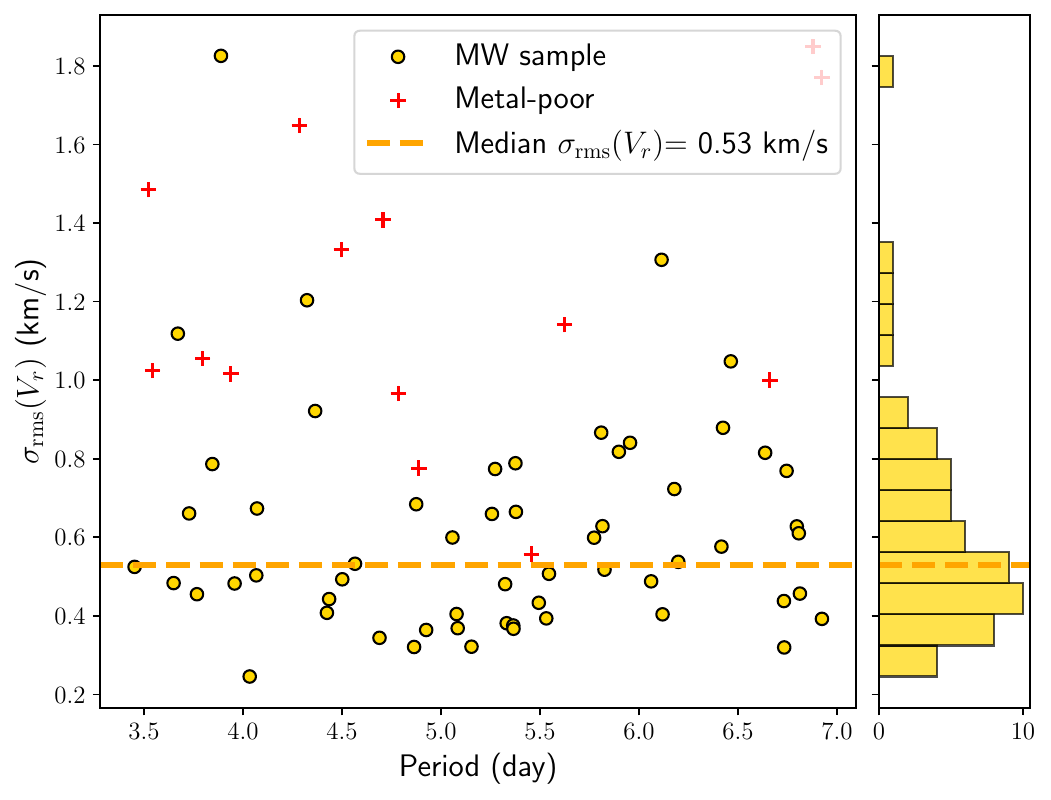}
    \caption{}
    \label{fig:residual_period}
    \end{subfigure}
    \hfill
    \begin{subfigure}[b]{0.45\textwidth}
    \includegraphics[width=\textwidth]{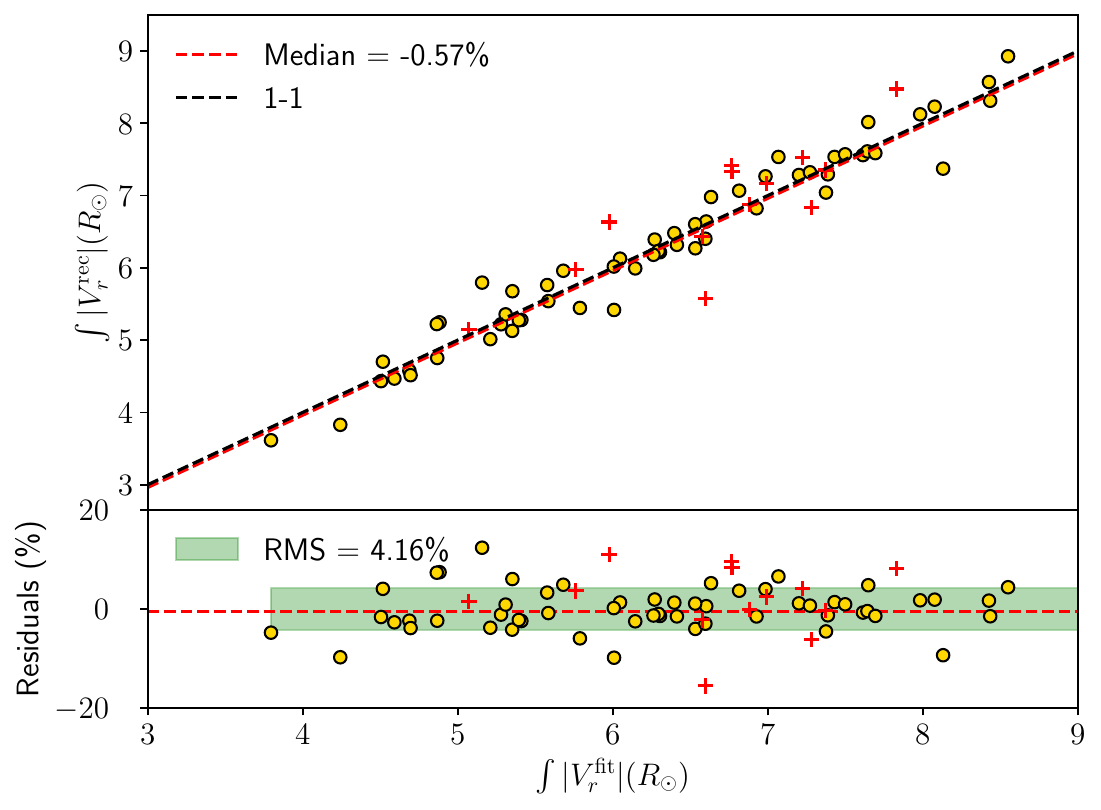}
    \caption{}
    \label{fig:residual_int_bis}
    \end{subfigure}
    
    \vskip\baselineskip
    
    \begin{subfigure}[b]{0.45\textwidth}
    \includegraphics[width=\textwidth]{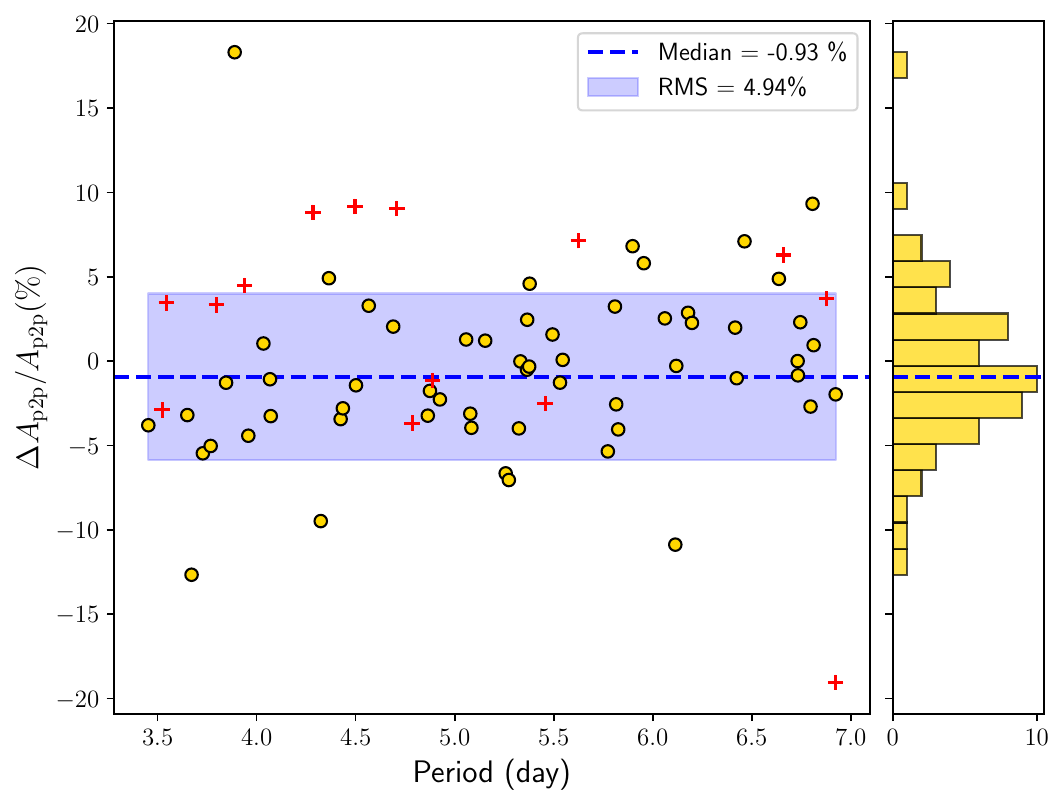}
    \caption{}
    \label{fig:delta_amp}
    \end{subfigure}
    \hfill
    \begin{subfigure}[b]{0.45\textwidth}
    \includegraphics[width=\textwidth]{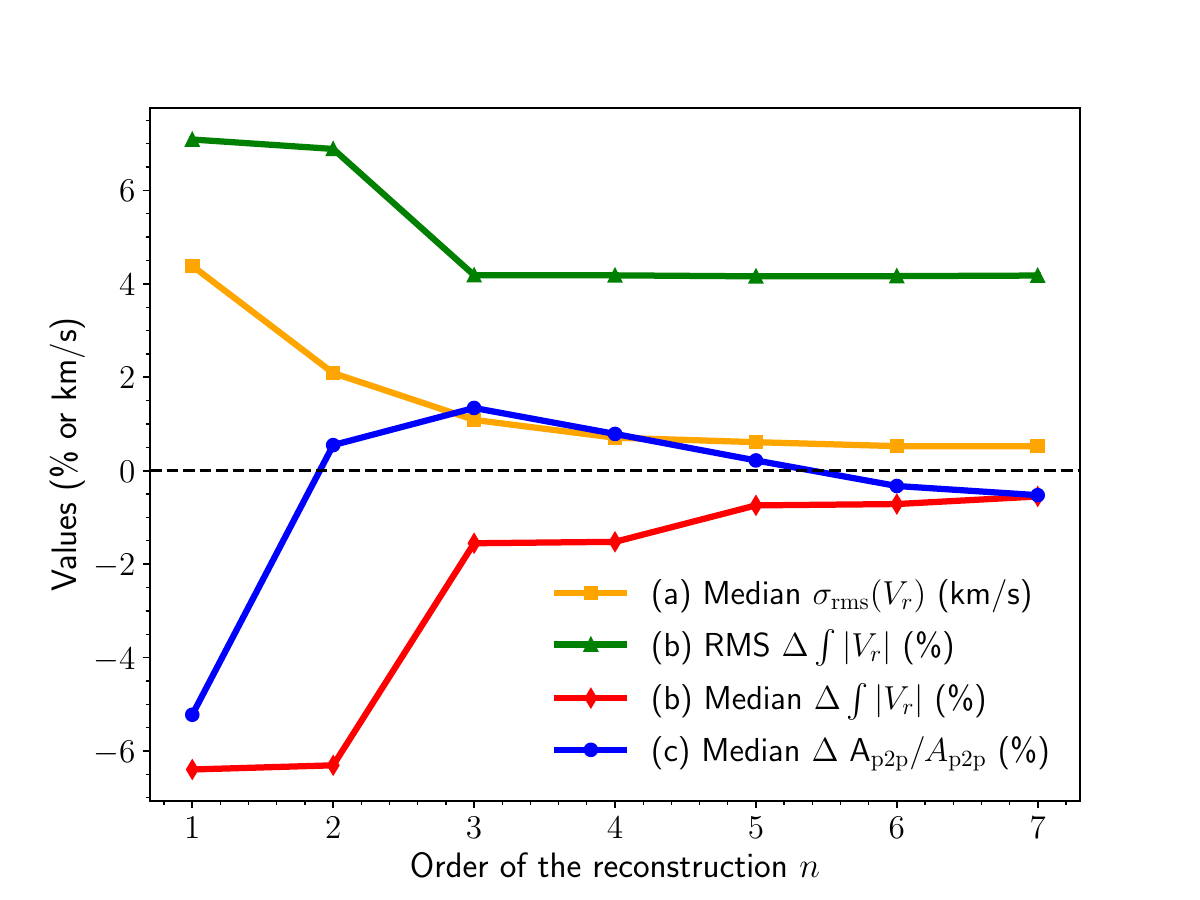}
    \caption{}
    \label{fig:optimal_order}
    \end{subfigure}
  
\caption{\small Internal accuracy of reconstructed RV curves for calibrating Cepheids and
   for metal-poor Cepheids (yellow circles and red crosses, respectively). (a) Reconstruction residuals, $\sigma_{rms}(V_r)$, for individual
   stars, as defined by Eq.~\ref{eq:sigma_vr}. The orange dashed line indicates the
   median of the residuals for the calibrating sample. (b) Comparison of absolute integrated RV curves, as defined in Eq.~\ref{eq:total_int}.  In the lower
   panel we plot the differences between the two values. The red dashed line indicates the median of the differences, with the
   green shaded area marking their RMS. (c) Relative peak-to-peak amplitude difference between the reconstructed RV curves and true RV curves (Fourier fits), as defined in  Eq.~\ref{eq:Ap2p}. The blue dashed line indicates the median value for the calibrating sample. (d) Accuracy of the RV curve reconstruction for Cepheids of the
   calibrating sample, vs. the Fourier order (see also Table~\ref{tab:fourier_order_rv}). The different lines represent the four previously discussed measures of accuracy,
   plotted with the same colors as in Figs.~9a,b,c (See Sect.~\ref{sect:order}).}
\label{fig:integrated}
\end{figure*}

The accuracy of the integrated RV curve is a key factor in the BW method \citep{nardetto04,Marengo2004}. In Fig.~\ref{fig:residual_int_bis}, we compare the integrated RV curves obtained with the reconstruction and the true ones, derived with the spectroscopic RV data (Fourier fits). To this end, we computed the integrated absolute value of the RV curves along the pulsation cycle, which is equivalent to the total linear radius variation $\Delta R$ normalized by the projection factor\footnote{The projection factor is defined as the ratio of the pulsational velocity and the measured radial velocity. This is a key parameter of the BW method (see \citealt{Trahin2021,Nardetto2023} and references therein).} $p$:
\begin{equation}\label{eq:total_int}
   \frac{\Delta R}{p} =\int_0^1 |V_r(\phi_i)|d\phi.
\end{equation}

The differences between $\Delta R/p$ computed from reconstructed RV curves and from the true RV curves (Fourier fits) has a standard deviation of 4.16\%. This error must be accounted for in the BW method, as it propagates directly into the error budget of the derived distances \citep{Marengo2004}. Nevertheless, the median difference of the total integrated RV curves remains below 1\% at $-0.57\%$. This result is crucial for application of the BW method based on RV curves reconstruction. Indeed, while individual stars may carry a few percent uncertainty, the integrated reconstructed RV curve averaged over a Cepheid population of 59 stars is accurate to better than 1\%. Somewhat surprisingly, the integrated RV curves of metal-poor Cepheids also show a good performance with a slightly larger dispersion of 6.96\% and a median difference of 2.00\% with respect to their Fourier fit. We deduce that the integrated RV curves of Cepheids are not very sensitive to the fit instabilities and high-order errors. We conclude that the integrated reconstructed RV curves of Cepheids, when used  in the context of the BW method over a sample of a large number of stars induce for solar-metallicity Cepheids (resp. metal-poor Cepheids), systematics no larger than 1\% (resp. 2.0\%), with statistical precision better than 5\% (resp. 7\%).

We also derive the relative difference of peak-to-peak amplitudes between the reconstructed RV curves and the true RV curves (Fourier fits)
\begin{equation}\label{eq:Ap2p}
    \Delta A_\mathrm{p2p}/A_\mathrm{p2p}=\frac{A_\mathrm{p2p}(V_r^\mathrm{rec})-A_\mathrm{p2p}(V_r^\mathrm{fit})}{A_\mathrm{p2p}(V_r^\mathrm{fit})}
\end{equation}

Within our sample, we find that the median of the relative difference of amplitudes is smaller than 1\% (see dashed blue line in Fig.~\ref{fig:delta_amp}) with a dispersion of about 5\% (see blue area in Fig.~\ref{fig:delta_amp}). We do not find significant trend within the sample. As in case of the previously discussed indicators, the metal-poor
Cepheids with low quality RV fits show higher dispersion (7.24\%) and higher median (3.57\%) of the relative amplitude difference, $\Delta A_\mathrm{p2p}/A_\mathrm{p2p}$

\subsection{Optimal order of reconstruction}\label{sect:order}
From the histogram of the order of the Fourier fits (Fig.~\ref{fig:order_histo}), we see that most of the RV curves are modeled with fits of order between 5 and 7, with a median of $n = 6$.
For pulsation periods below 7\,days, 48 out of 59 stars of our calibrating sample (i.e. 83\%) have Fourier fits of order $n\leq7$, while fits of order between 5 and 7 are used in 42 stars (71\%). These results are comparable with the median order of the fit determined for the short-period Cepheid sample ($n=7$, 47 stars) of \cite{Anderson2024}. However, the order of the fit is sensitive to the stellar properties such as the amplitude and the pulsation period, but also to the quality of the data such as the accuracy of the measurements, number of data points and the phase coverage. 

To reconstruct the shape of RV curve of a Cepheid, we first have to decide what Fourier order should be used.
Unfortunately, we did not find a way to determine the order of the RV fit, for example as a function of the order of the LC fit and the pulsation period. This is so because of a poor correlation between the fit order of the RV curves and of the LCs, as we can see in Fig.~\ref{fig:2d_histo}. Instead, we chose to find the order that will be optimal for reconstructing the RV curves for all Cepheids of the calibrating sample. To this end, we reconstructed the RV curves of the calibrating sample using different orders from $n=1$ to $n=7$, and then for each reconstruction we computed different measures of accuracy:
\begin{itemize}
    \item the median of RV reconstruction residuals, $\sigma_{rms}(V_r)$ as defined by Eq.~\ref{eq:sigma_vr} and plotted in Fig.~\ref{fig:residual_period}
    \item the RMS of differences and the median bias of the absolute integrated RV curves, as plotted in Fig.~\ref{fig:residual_int_bis} 
    \item and the median of relative amplitude difference, $\Delta A_\mathrm{p2p}/A_\mathrm{p2p}$ as presented in Fig.~\ref{fig:delta_amp}.
\end{itemize}
 We plot the result in Fig.~\ref{fig:optimal_order}, see also Table \ref{tab:fourier_order_rv}. Overall, the quality of the reconstruction improves rapidly up to order 3, and the addition of higher orders has only a small, yet non-negligible, effect on the accuracy of the reconstruction. We find that a fit of order of n=5-7 is optimal for reconstruction of RV curves of the calibrating sample, because it minimizes all four measures of accuracy. We decide to use n=6 as our final choice. This is the same values as the median order of the fit for the calibrating sample.

Therefore, our reconstruction overestimates the fit order of about one third of the calibrating sample (order strictly below 6, 19 stars out of 59) and underestimates the fit order for 42\% of the sample (25 out of 59). Neglecting orders higher than 6 has a negligible impact on the reconstruction quality, because the amplitudes of higher orders harmonics constitute less than 5\% of the amplitude of the first harmonic (see for example $R_{71}$ in Fig.~\ref{fig:R71}).

However, the error introduced could be larger in the case of RV curves with low fit order.
This effect is largely mitigated by the fact that only 10\% of the calibrating sample have fit order strictly below 5, so the number of Cepheids with potentially larger error of the fit remains limited. Besides, the Cepheids with an order of the fit of 4 or less, namely CF~Cas ($\sigma_{rms}=0.68$\,km/s), V1154~Cyg ($\sigma_{rms}=0.36$\,km/s), V1162~Aql ($\sigma_{rms}=0.79$\,km/s), V733~Aql ($\sigma_{rms}=0.72$\,km/s) and V496~Aql ($\sigma_{rms}=0.61$\,km/s), usually do not have significant errors, except SS~Sct ($\sigma_{rms}=1.12$\,km/s). This can be explained by the fact that RV curves modeled with low-order Fourier fits  usually have lower amplitudes and, consequently, lower amplitude ratios. Therefore, higher Fourier orders contribute less to the overall error of the reconstruction. In conclusion, while individual determination of the fit order is desirable for the most optimal reconstruction of individual stars, we are confident that fixing the order of the fit at $n=6$ is sufficient for most of the Cepheids.

\begin{table}
\centering
\caption{Accuracy of the reconstructed RV curves shape as a function of the Fourier fit order.}
\label{tab:fourier_order_rv}
\begin{tabular}{c c c c c}
\hline
$n$ & $\Delta A/ A$& $\sigma_\mathrm{RMS}$& RMS $\Delta \int |V_r|$&  Median $\Delta \int |V_r|$\\
 & (\%) & (km/s) &  (\%) & (\%)\\
\hline
1 & $-5.223$ & $4.384$ & $7.088$ & $-6.395$ \\
2 & $0.548$  & $2.090$ & $6.886$ & $-6.305$ \\
3 & $1.343$  & $1.087$ & $4.184$ & $-1.553$ \\
4 & $0.787$  & $0.706$ & $4.181$ & $-1.522$ \\
5 & $0.220$  & $0.611$ & $4.160$ & $-0.742$ \\
6 & $-0.327$ & $0.526$ & $4.164$ & $-0.714$ \\
7 & $-0.524$ & $0.526$ & $4.176$ & $-0.552$ \\
\hline
\end{tabular}
\begin{tablenotes}
\footnotesize
\item \textbf{Note.} The table presents the relative amplitude (see Eq.~\ref{eq:Ap2p}), the residuals of the reconstruction (see Eq.~\ref{eq:sigma_vr}).  The last column gives the median relative deviation of the reconstructed RVs in percent. See also Fig.~\ref{fig:optimal_order}.
\end{tablenotes}
\end{table}

\subsection{Monte Carlo Cross-validation}\label{sect:montecarlo}
\begin{figure*}[htbp]
    \centering
    \begin{subfigure}[b]{0.33\textwidth}
        \includegraphics[width=\linewidth]{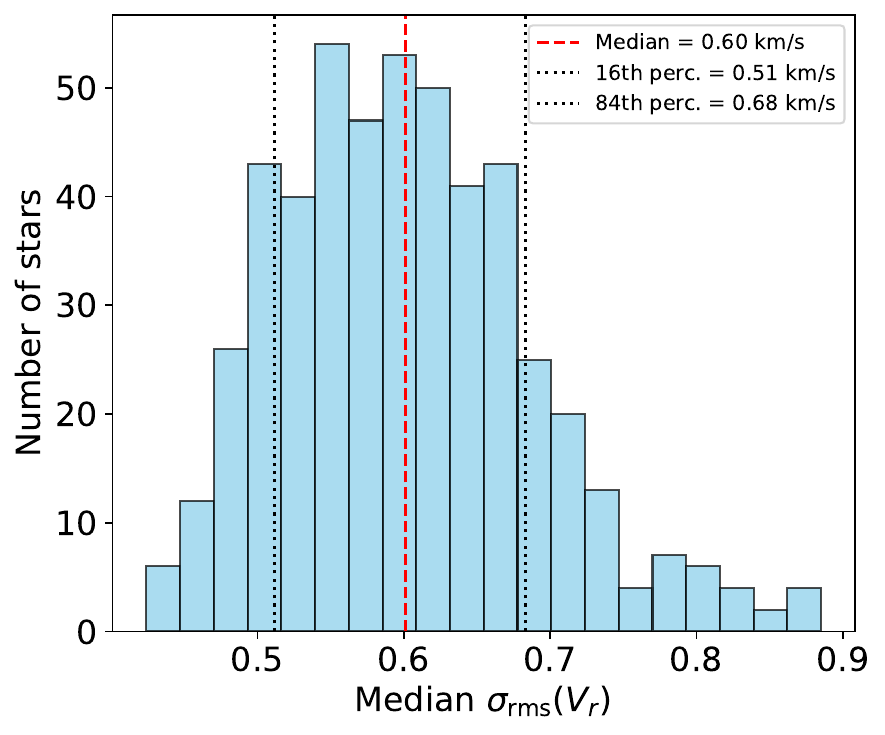}
        \caption{}
        \label{fig:mc_res}
    \end{subfigure}
    \begin{subfigure}[b]{0.33\textwidth}
        \includegraphics[width=\linewidth]{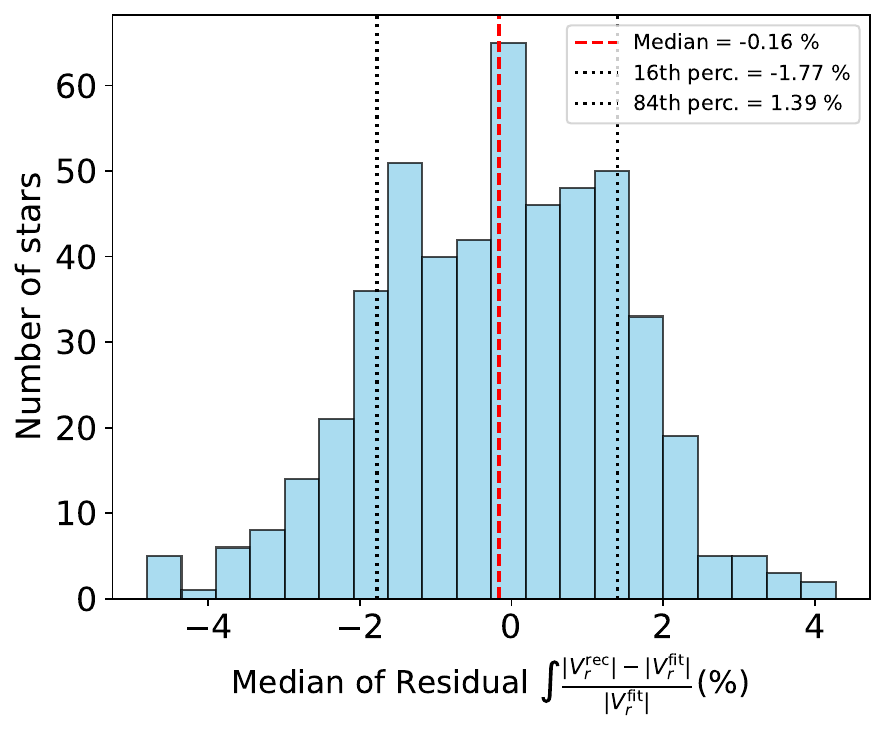}
        \caption{}
        \label{fig:mc_int}
    \end{subfigure}
    \begin{subfigure}[b]{0.33\textwidth}
        \includegraphics[width=\linewidth]{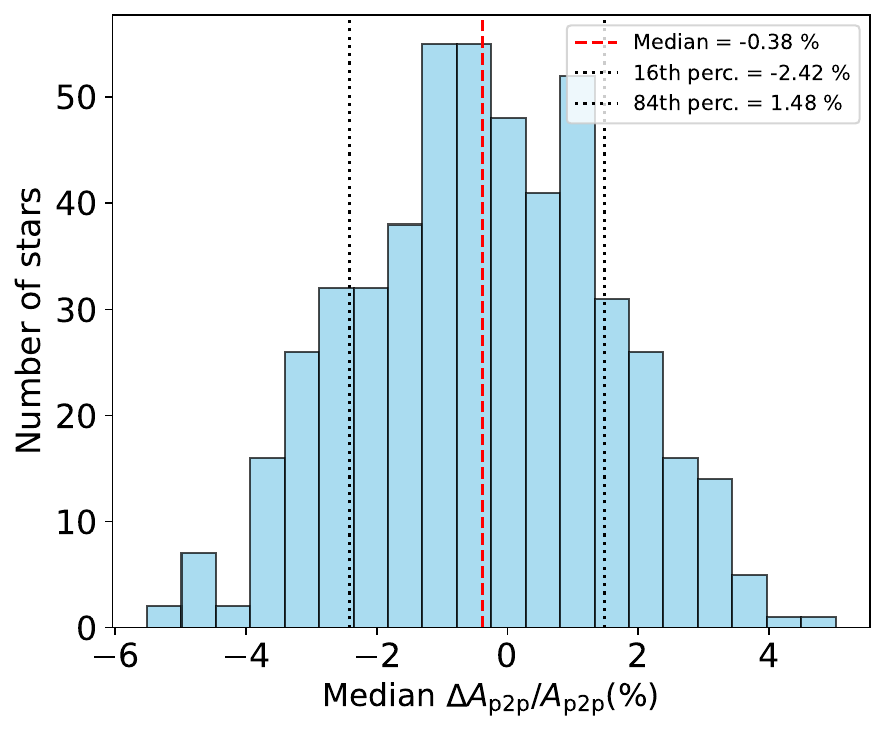}
        \caption{}
        \label{fig:mc_amp}
    \end{subfigure}
\caption{\small External accuracy of the reconstructed RV curves from Monte Carlo resampling (see Sect.~\ref{sect:montecarlo}). In each plot, the red dashed line indicates the median of the distribution. (a) Reconstruction residuals, $\sigma_\mathrm{rms}(V_r)$ as defined by Eq.~\ref{eq:sigma_vr}. (b) Difference between the absolute integrated RV curves (see Eq.~\ref{eq:total_int}), computed from the reconstructed and from the true RV data. (c) Relative peak-to-peak amplitude difference between the reconstructed RV curves and true RV curves (Fourier fits), as defined in Eq.~\ref{eq:Ap2p}.}
    \label{fig:montecarlo}
\end{figure*}

To assess realistic external accuracy and the generalization capability of our method, we applied a bootstrapping approach. In each realization, we randomly select 20\% of the stars from the calibrating sample (12 stars), while the remaining 47 stars form the training set. This procedure is repeated 500 times with different random realizations of the test sample. External accuracies are derived from the reconstruction residuals obtained for each test subset, following the same procedure as for the previous measurements. The median of reconstruction residuals provides a realistic estimate of the overall performance and predictive power of our method. The result of the Monte Carlo procedure is presented in Fig.~\ref{fig:montecarlo}. We obtain results which are consistent with the overall accuracy of our calibration. We show that our method is able to reconstruct external RV curves with a median accuracy of 0.60\,km/s as compared to the Fourier fits of true spectroscopic RV measurements (see Fig.~\ref{fig:mc_res}). The absolute integrated RV curves are also accurate, with a median difference of only $-0.16$ with respect to the true RV curves (see Fig.~\ref{fig:mc_int}). 

\section{Template fitting method}\label{sect:template_fitting}
\begin{figure*}[!htbp]
\centering
\begin{subfigure}[b]{.35\textwidth}
  \includegraphics[width=\linewidth]{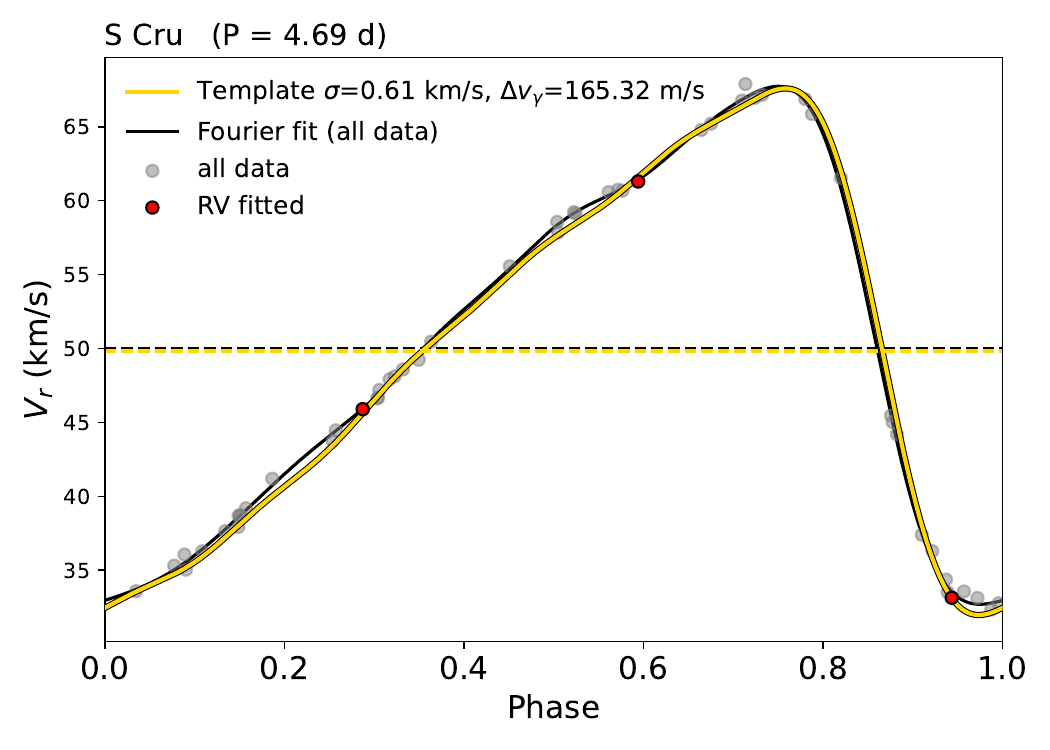}
  \caption{}
  \label{fig:template_s_cru}
\end{subfigure}%
\begin{subfigure}[b]{.35\textwidth}
  \includegraphics[width=\linewidth]{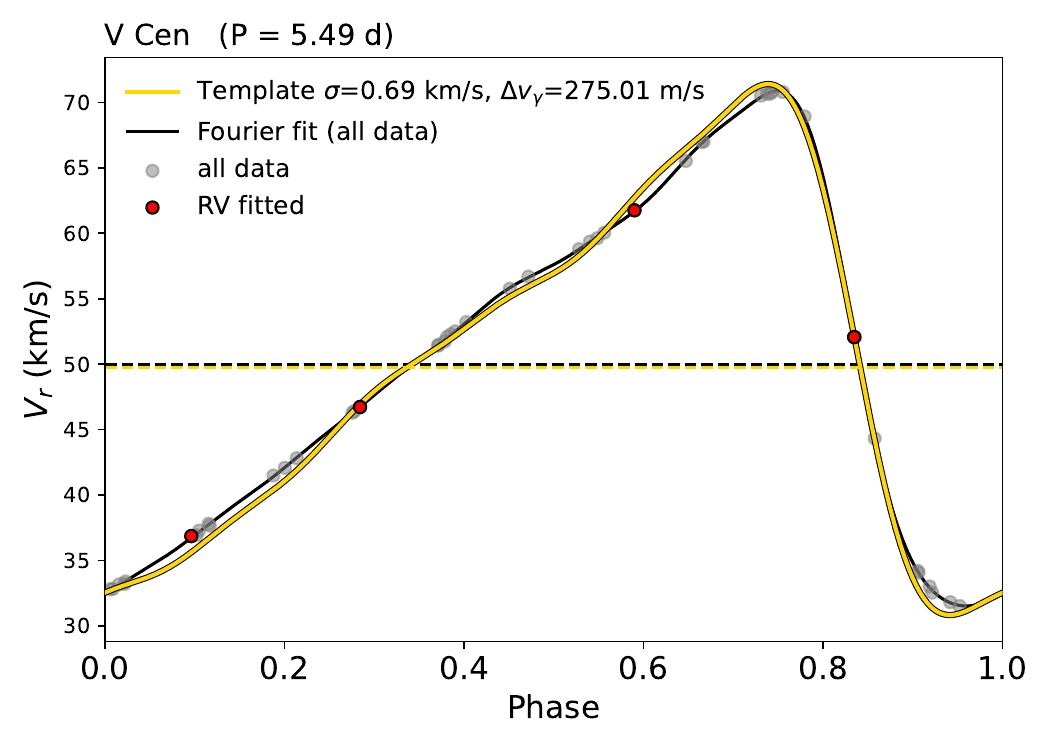}
  \caption{}
  \label{fig:template_v_cen}
\end{subfigure}%
\begin{subfigure}[b]{.35\textwidth}
  \includegraphics[width=\linewidth]{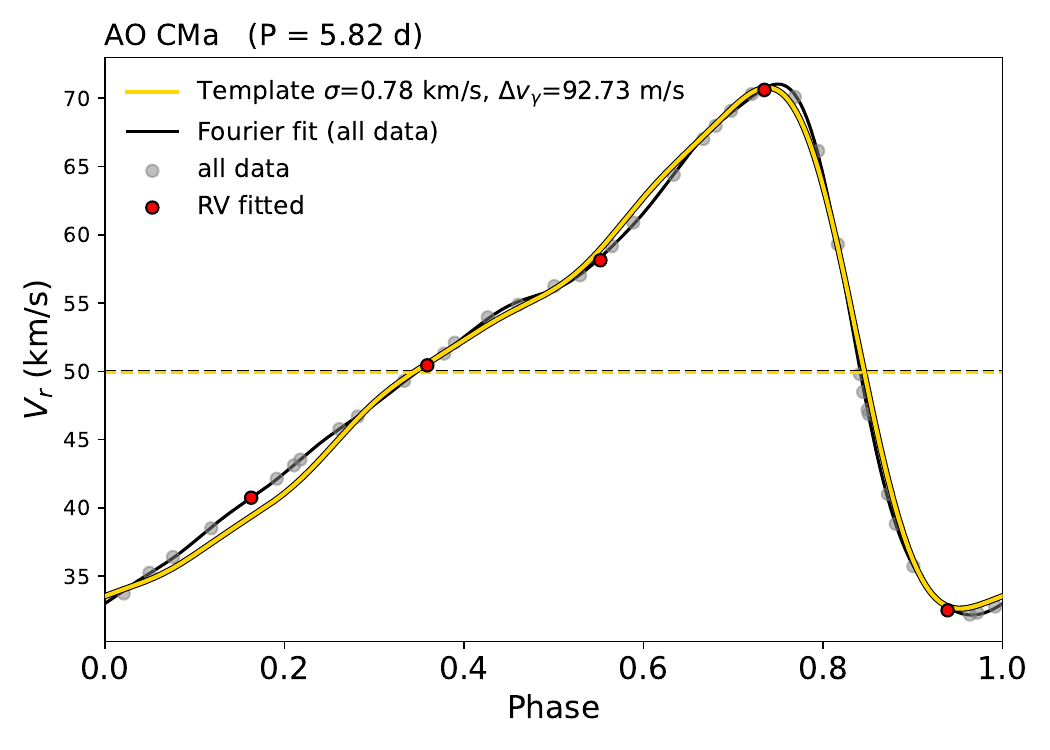}
  \caption{}
  \label{fig:template_ao_cma}
\end{subfigure}\vskip\baselineskip
\begin{subfigure}[b]{.35\textwidth}
  \includegraphics[width=\linewidth]{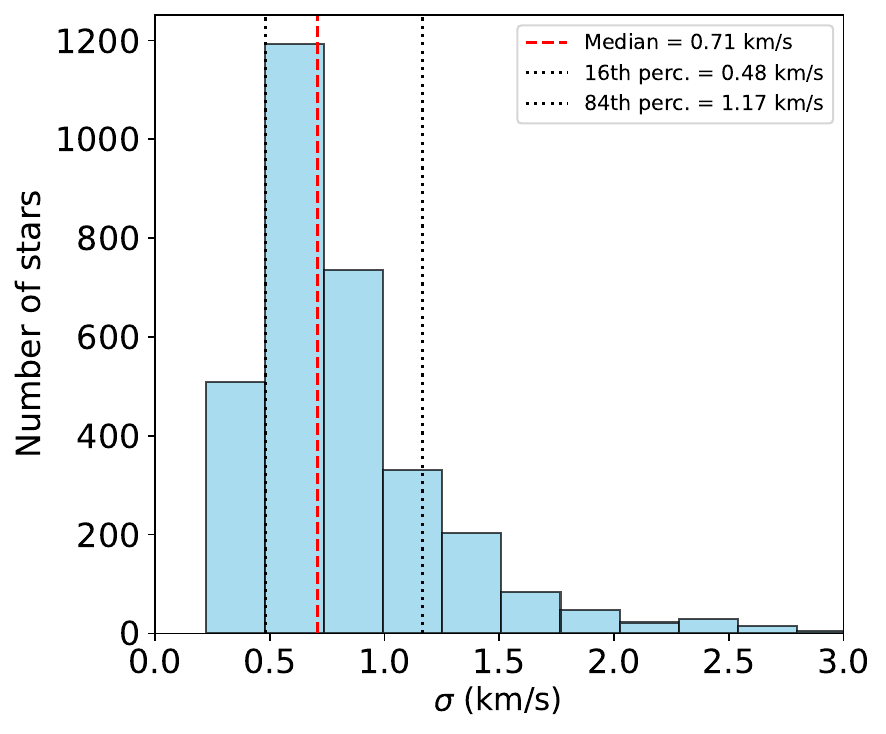}
  \caption{}
  \label{fig:rms_3}
\end{subfigure}%
\begin{subfigure}[b]{.35\textwidth}
  \includegraphics[width=\linewidth]{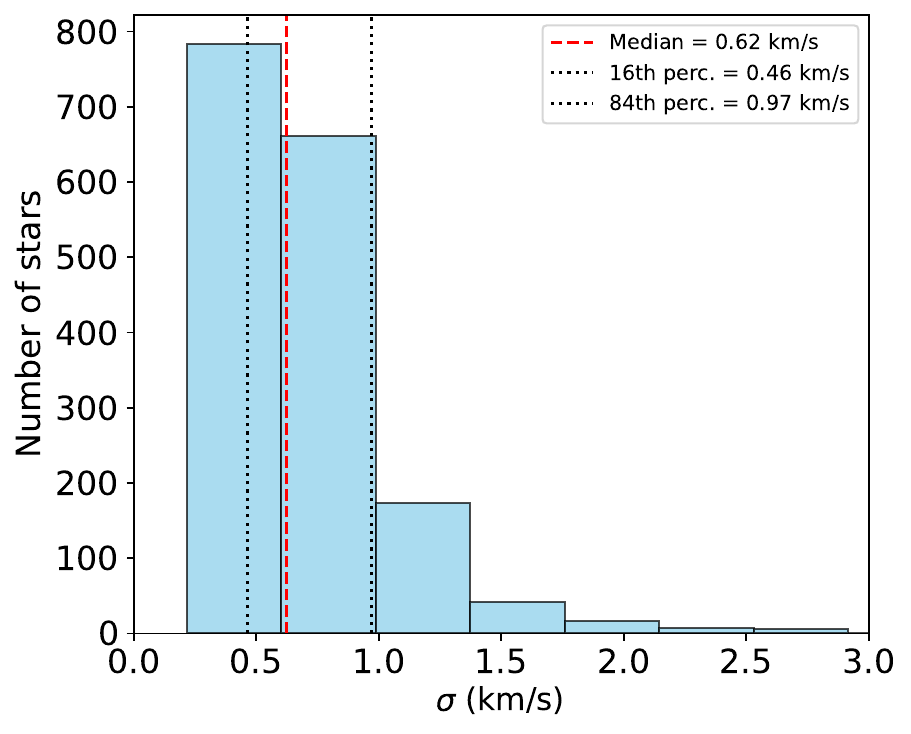}
  \caption{}
  \label{fig:rms_4}
\end{subfigure}%
\begin{subfigure}[b]{.35\textwidth}
  \includegraphics[width=\linewidth]{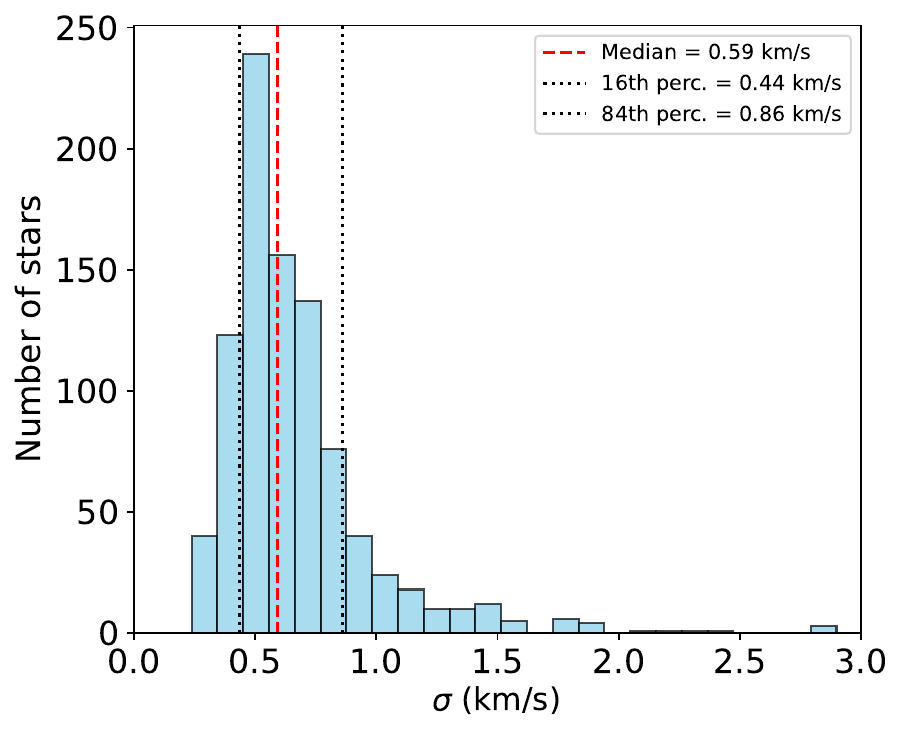}
  \caption{}
  \label{fig:rms_5}
\end{subfigure}\vskip\baselineskip
\begin{subfigure}[b]{.35\textwidth}
  \includegraphics[width=\linewidth]{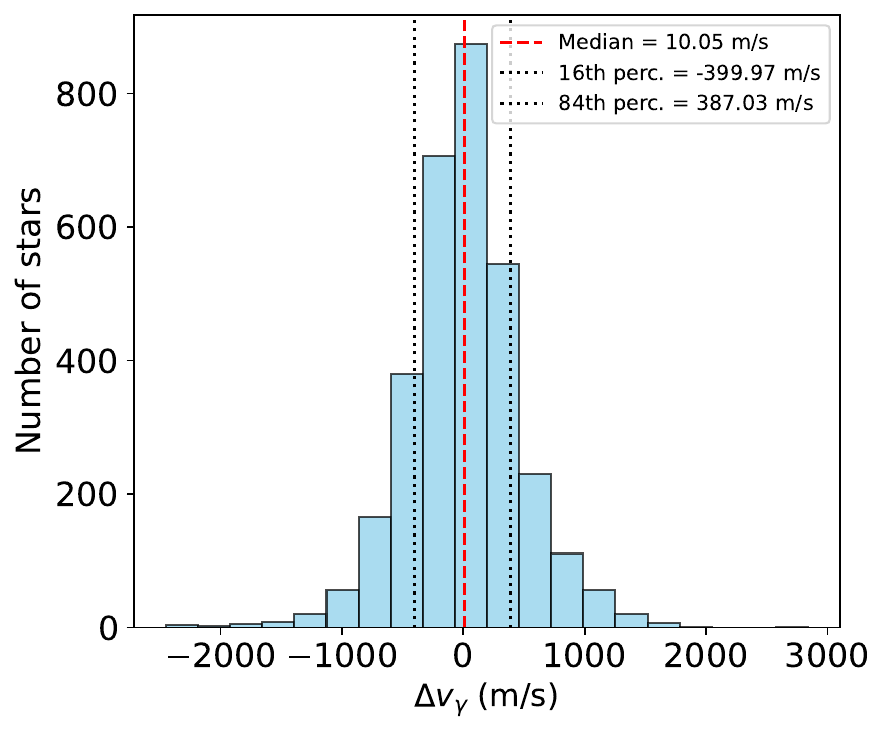}
  \caption{}
  \label{fig:vgamma_3}
\end{subfigure}%
\begin{subfigure}[b]{.35\textwidth}
  \includegraphics[width=\linewidth]{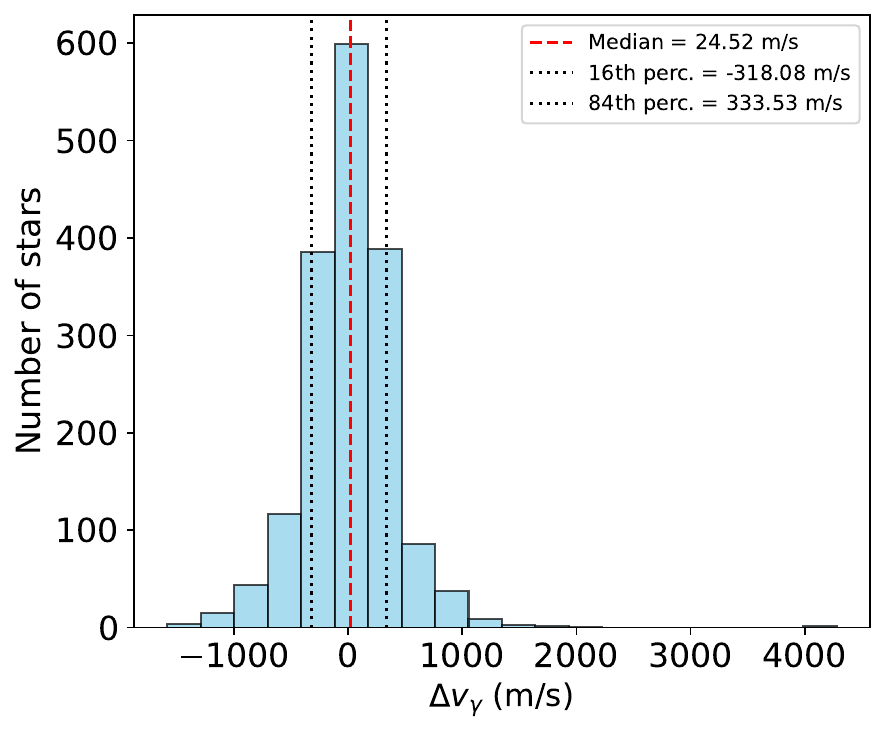}
  \caption{}
  \label{fig:vgamma_4}
\end{subfigure}%
\begin{subfigure}[b]{.35\textwidth}
  \includegraphics[width=\linewidth]{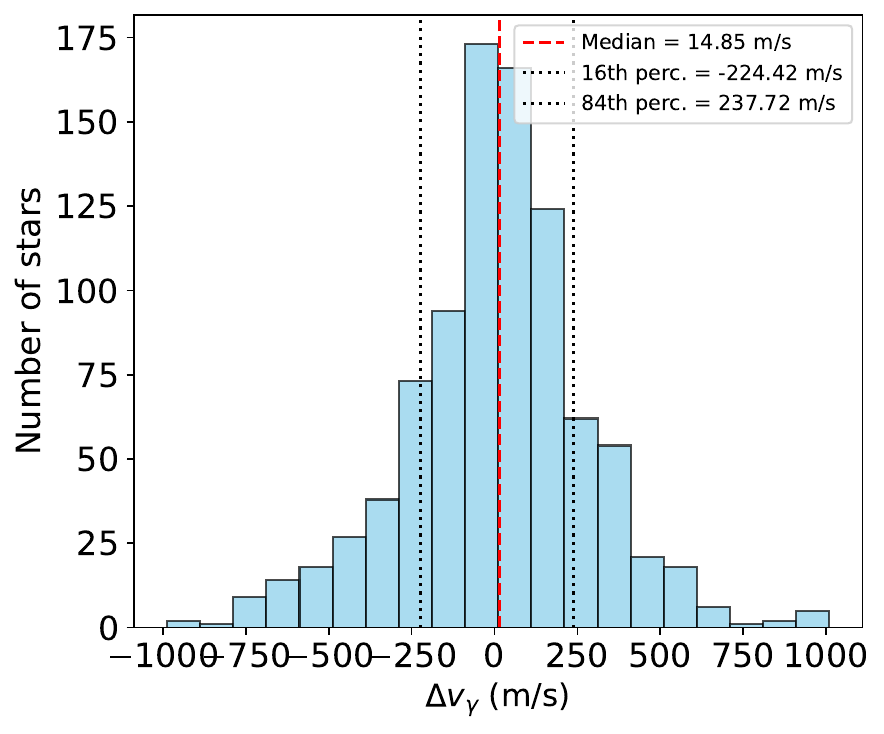}
  \caption{}
  \label{fig:vgamma_5}
\end{subfigure}\vskip\baselineskip

\caption{Template fitting accuracy for 3,4 and 5 RV measurements in each column, respectively. The first row (a) to (c) displays example of template fitting for three different stars. The second row (d) to (f) presents template accuracy with respect to the Fourier fit (true RV curve) obtained from bootstrapping method on the test sets. The last row presents the difference in systemic velocity from each template fitting with respect to the true value, also presented in Table \ref{tab:Nrv_sigma_dvgamma}.
}
\label{fig:template_accuracy}
\end{figure*}

\begin{figure}[htbp]
    \centering
    \begin{subfigure}[b]{0.24\textwidth}
        \includegraphics[width=\linewidth]{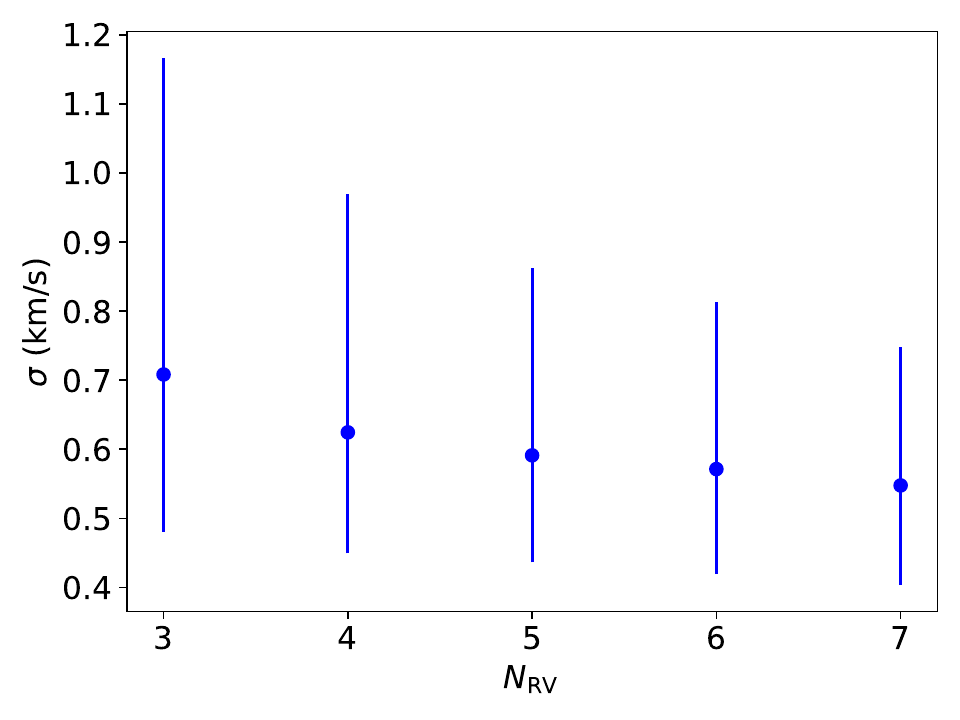}
        \caption{}
        \label{fig:sig_RV}
    \end{subfigure}
    \begin{subfigure}[b]{0.24\textwidth}
        \includegraphics[width=\linewidth]{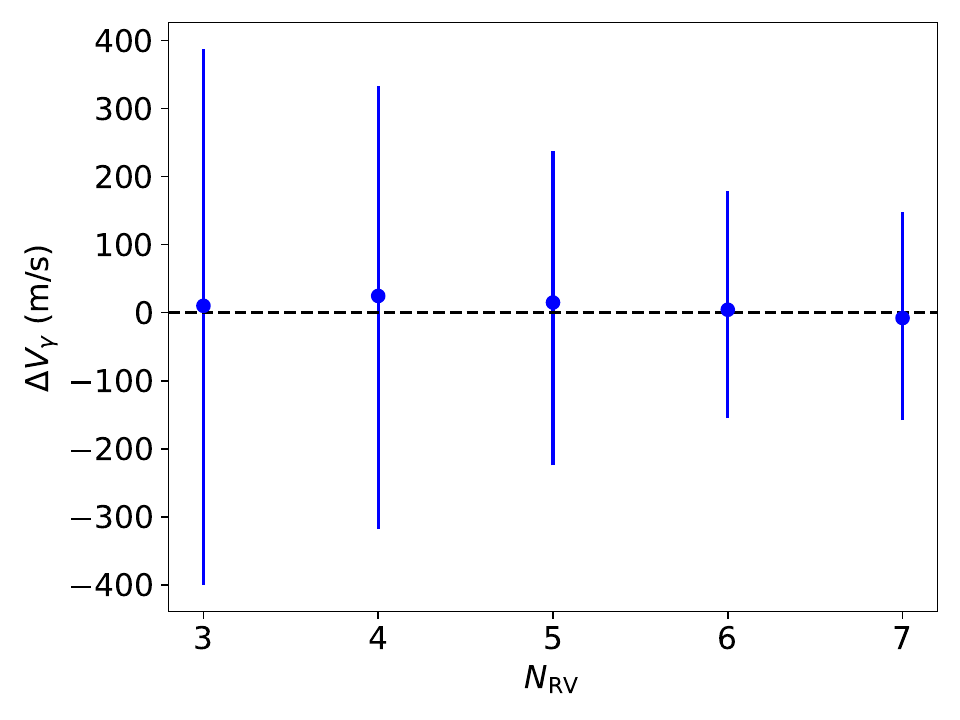}
        \caption{}
        \label{fig:gamma}
    \end{subfigure}
    \caption{Accuracy of the template fitting method as a function of the number of RV measurements. (a) RMS of the residual between fitting and template and the Fourier fit of the true RV curves. (b) Difference of systemic velocity.}
    \label{fig:Nrv_vs_sigma}
\end{figure}

In the previous section, we demonstrated the ability of our method to reconstruct precisely the shape of the RV curves of Cepheids. We now estimate the accuracy of this technique as a template fitting method. Indeed, given few RV measurements available for a Cepheid, this is possible to adjust the phase and the systemic velocity in order to fit the RV model obtained from the shape of the LC. In order to test these templates, we used the same train and test samples generated for the bootstrapping method presented in Sect.~\ref{sect:montecarlo}. For each star of the validation sample, we selected 3,4,5,6 and 7 RV measurements from the initial data set presented in Sect.~\ref{sect:Source_RV}. These RV measurements were selected to be uniformly distributed along the pulsation cycle, starting from a random pulsation phase for each star. This procedure allows to avoid as much as possible any colinear points that are not able to constrain the template fitting. While spectroscopic survey may not ensure a regular cadence of observation, our method ensures robust accuracy determination in optimal conditions. For each sample of RV measurements, we arbitrarily offset the phase and the systemic velocity. We then use the RV model derived from the shape of the LCs as determined by the training set, and we applied a least-square method to adjust the phase and $v_\gamma$ in order to fit the RV measurements. The results for 3,4 and 5 RV measurements are presented in Fig.~\ref{fig:template_accuracy} and for 3 to 7 RV measurements in Table \ref{tab:Nrv_sigma_dvgamma} and Fig.~\ref{fig:Nrv_vs_sigma}.

Our results show that with only 3 RV measurements regularly spaced along the pulsation cycle, our templates are capable of fitting the RV measurements with a median uncertainty of 0.7\,km/s with respect to the true Fourier fit on the available RV data. This accuracy improves with the number of RV measurements. The systemic velocity is also determined accurately, to better than 30\,m/s, with a typical precision of a few 100\,m/s, depending on the number of RV measurements used to fit the templates.

\begin{table}
\centering
\caption{Accuracy of the template fitting.}
\label{tab:Nrv_sigma_dvgamma}
\begin{tabular}{c c c}
\hline
$N_{\rm RV}$ & $\sigma$\,(km/s) & $\Delta V_\gamma$\,(m/s)\\
\hline
3 & $0.71^{+0.46}_{-0.23}$ & $+10^{+377}_{-410}$ \\
4 & $0.62^{+0.35}_{-0.18}$ & $+25^{+309}_{-343}$ \\
5 & $0.59^{+0.27}_{-0.16}$ & $+15^{+223}_{-239}$ \\
6 & $0.57^{+0.24}_{-0.15}$ & $+4^{+174}_{-159}$ \\
7 & $0.55^{+0.20}_{-0.14}$ & $-8^{+156}_{-150}$ \\
\hline
\end{tabular}
\begin{tablenotes}
\footnotesize
\item \textbf{Note.} Reconstruction residuals ($\sigma$) as defined by Eq.~\ref{eq:sigma_vr} and systemic velocity shift ($\Delta V_\gamma$) as a function of the number of RV measurements $N_{\rm RV}$. Quoted values correspond to the median with the 16th and 84th percentiles (see also Fig.~\ref{fig:template_accuracy}).
\end{tablenotes}
\end{table}

\section{Discussion}\label{sect:discussion}

\subsection{Calibration of the phase lag between LC and RV curves}
\begin{figure}[]
    \includegraphics[width=0.45\textwidth]{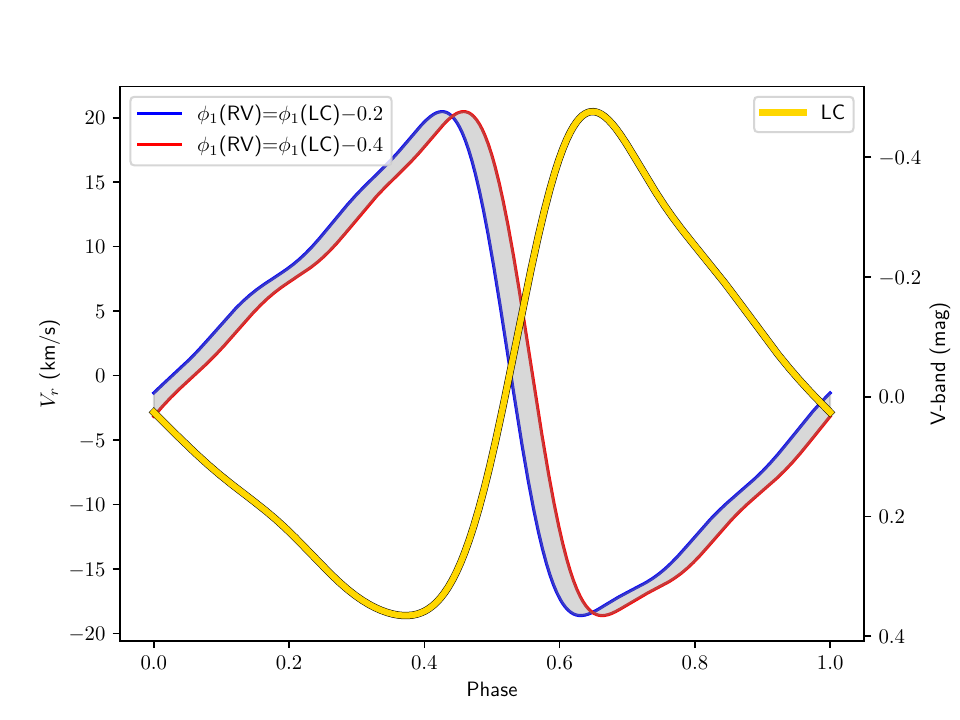}
    \caption{LC and reconstructed RV curve of $\delta$ Cep using phase lag range derived by \cite{Ogloza2000a}. The grey zone represents the possible location of the RV curve.}
    \label{fig:phi1}
\end{figure}

In this paper, we proposed a method to reconstruct the shape of the RV curves. However, we are not yet able to phase accurately the RV curves with respect to the LC, as we have no information on the phase difference between the two. In order to reconstruct the RV curves with proper phases, it is necessary to derive this phase difference, also called the phase lag, defined as $\Delta \phi_1=\phi_1(RV)-\phi_1(LC)$, with $\phi_1$ being the phase of the first harmonic. The values of phase lag were measured for a sample of Galactic Cepheids by \cite{Ogloza2000a}, who found a mean value of $\Delta\phi_1 =-0.28\pm 0.10$ rad, independent of the pulsation period. for a sample of Galactic Cepheids. To illustrate the range of values for $\Delta \phi_1$, we displayed in Fig.~\ref{fig:phi1} an example of RV curve reconstruction for $\delta$ Cep, phased with respect to the LC using $\Delta \phi_1=-0.28\pm0.10$. As we can see, an uncertainty of $\Delta \phi_1$ of only 0.2\,rad translates to an uncertainty of the pulsation
phase of only $\delta\phi$ = 0.03. This narrow range can be useful to constrain the phasing of the reconstructed RV curves for application of the BW method.

However, a note of caution is necessary. \cite{Szabo2007} showed theoretically that the phase lag depends on the Mass-Luminosity relation and possibly also weakly on the metallicity of the Cepheids. Therefore, while the
range of $\Delta\phi_1$ determined by \cite{Ogloza2000a} can in principle be used for constraining relative phases of RV curves and LCs, a more precise calibration is needed, focusing on the pulsation period range of our 
method. We will address this topic in a paper in preparation.

\subsection{Limitations and improvements}
The method presented in this paper shows the possibility to reconstruct the RV curves of Cepheids with reasonable precision. Several aspects can be however improved in the future in order to perform the reconstruction with even higher precision. We showed in this paper the good agreement of the metal-poor Cepheids with our calibrating sample, but only precise measurements will make it possible to investigate the metal dependency in more detail. Thus it is important to increase the effort of spectroscopic observations of metal-poor Cepheids, particularly with $P\sim4\,$days, which exhibit much larger amplitudes and amplitude ratios. RV measurements utilizing the Calcium triplet of LMC and SMC Cepheids from Gaia DR4 will be also useful for comparison with our reconstructed RV curves.

Another source of uncertainty is a possible secular or irregular period change during the Cepheid evolution \citep{Csornyei2022,Rathour2025}. Since our calibrating sample consists of RV curves and LCs that were not obtained at the same epoch, a part of the dispersion of the empirical relations between their Fourier parameters might be caused by this effect. Therefore, simultaneous observations of both the LC and the RV curves would certainly improve the calibration.

Extending the RV reconstruction method to other photometric bands will be important in order to apply the method to different photometric surveys. It is particularly relevant to calibrate our empirical relations directly into the Sloan bands, which are used by the Vera Rubin Telescope.  Moreover, while our work focuses only on short-period Cepheids between 3 and 7 days, extending the method to long-period Cepheids could be useful for reconstructing RV curves of more distant extragalactic Cepheids. Our method could be also calibrated using \textit{Gaia} LCs and RVs.

\subsection{Toward extrinsic RV curve templates}
Up to now, templates for RV curves remain intrinsic, i.e. Fourier fit for a given star is used to fit few RV measurements obtained at a different epoch. This method is used for the determination of the systemic velocity induced by the orbital motion \citep{Anderson2016RV,Anderson2019,Nardetto2024,Shetye2024}. On the contrary, there are no extrinsic template fitting methods allowing to determine the mean velocity of Cepheids without a priori knowledge of their RV curves. Such templates are already mature in the case of LCs \citep{Inno2015,Bras2025}, allowing to determine the mean magnitude of extragalactic Cepheids in different bands. Our reconstruction method, on the other hand, could be useful for this purpose for any short-period Cepheids for which the $V$-band LC is available. This opens the possibility of detecting orbital motion for a larger number of Cepheids, while reducing the number of required spectroscopic (RV) observations. Alternatively, our reconstructions could be used for template-fitting to complement high-resolution spectroscopic observations in the context of the BW method.

\subsection{Toward a photometric BW method}
Apart from the systematic errors linked to the projection factor \cite{Trahin2021,Nardetto2023}, the application of the BW method remains limited by the difficulty of obtaining high-precision RV measurements for a statistically significant sample of extragalactic Cepheids. New generation of spectroscopic instruments ANDES and MOSAIC of the ELT will enable to obtain RV curves of Cepheids in the Local Group \citep{Ande2024,ANDES2024,MOSAIC2024}. However the RV curves will be either expensive to cover the pulsation cycle with limited telescope time, or beyond the capability of those instruments which will be limited to $m_{AB}=20\,$mag for integration times of $\sim1$\,hr.

In contrast, the method presented in this paper is easily applicable to thousands of Cepheids from galaxies of the Local Group thanks to their LCs. In combination with angular diameter measurements from precisely calibrated Surface Brightness Color Relations \citep{Bailleul2025}, this opens the possibility for a new purely photometric BW method. This is particularly relevant in the era of photometric surveys, such as the Vera Rubin Telescope which will be able to obtain excellent phase coverage for Cepheids down to $P=4\,$days, located in galaxies as distant as 4.4\,Mpc \citep{Hoffman2015}. In principle, well covered LCs with a minimum of 30-35 data points are enough to obtain accurate estimations of their pulsation periods and their Fourier parameters $R_{21}$ and $R_{31}$, which are the only information necessary to reconstruct the shape of the RV curves. Besides, our method is almost insensitive to blending and to the metallicity of Cepheids. In our next paper (in preparation), we show the potential to determine LMC and SMC distances and evaluate the level of systematics of a purely photometric BW method.

\section{Conclusion}\label{sect:conclusion}
In this paper, we calibrated a new method allowing the reconstruction of the shape of the RV curves of fundamental-mode Classical Cepheids, using only the pulsation periods and the Fourier parameters $R_{21}$ and $R_{31}$ of their V-band LCs. This new approach to the RV curves reconstruction is useful for understanding the relation between LC and RV curves, which in turn can help to constrain Cepheid hydrodynamical models. We show that our method is able to reconstruct the shape of RV curves with a precision of 0.60 km/s compared to the true RV curves determined with the spectroscopic measurements. For the sample of Cepheids, the integrated reconstructed RV curves (which are equivalent to $\Delta R/p$) are accurate on average to better than 1\% and precise to better than 5\%. The high accuracy and precision of the reconstruction makes the method ideal to apply it to a BW analysis of a statistically large sample of Cepheids. Future work will be useful to reduce the scatter of the empirical calibrations, to establish the calibrations in different photometric bands and to extend the range of pulsation periods.

\section{Data Availability}
Full Tables of Fourier parameters \ref{tab:fourier_lc} and \ref{tab:fourier_rv} are only available at the CDS via anonymous ftp to cdsarc.u-strasbg.fr (130.79.128.5) or via \url{http://cdsarc.u-strasbg.fr/viz-bin/cat/J/A+A/}

\begin{acknowledgements} The authors acknowledge the support of the French Agence Nationale de la Recherche (ANR), under grant ANR-23-CE31-0009-01 (Unlock-pfactor). B.P. gratefully acknowledges financial support from the Polish National
Science Centre grant SONATA BIS 2020/38/E/ST9/00486. R.S. acknowledges support from the SONATA BIS grant, 2018/30/E/ST9/00598, from the National Science Center, Poland. We also acknowledge support from the Polish Ministry of Science and Higher Education grant 2024/WK/02. The research leading to
these results has received funding from the European Research Council (ERC)
under the European Union’s Horizon 2020 research and innovation program
(projects CepBin, grant agreement 695099, and UniverScale, grant agreement
951549). AG acknowledges the support of the Agencia Nacional de Investigación Científica y Desarrollo (ANID) through the FONDECYT Regular grant 1241073. We acknowledge with thanks the variable star observations from the AAVSO International Database contributed by observers worldwide and used in this research. This research made use of the SIMBAD and VIZIER databases at CDS, Strasbourg (France) and the electronic bibliography maintained by the NASA/ADS system. This research also made use of Astropy, a community-developed core Python package for Astronomy \citep{astropy2018,astropy2022}. 
\end{acknowledgements}

\bibliographystyle{aa}  
\bibliography{bibtex_vh} 

\clearpage
\onecolumn

\begin{appendix}

\section{Data sets}
\begin{longtable}{l l | S[table-format=2.6] c | S[table-format=2.6] c}
\caption{Calibrating sample  \label{tab:data_ref}}
 \\
\hline\hline
Name & Quality & \multicolumn{2}{c}{RV curve} & \multicolumn{2}{c}{LC} \\
     &         & \multicolumn{1}{c}{Period} & Ref. & \multicolumn{1}{c}{Period} & Ref. \\
\hline
\endfirsthead
\hline\hline
Name & Quality & \multicolumn{2}{c}{RV curve} & \multicolumn{2}{c}{LC} \\
     &         & \multicolumn{1}{c}{Period} & Ref. & \multicolumn{1}{c}{Period} & Ref. \\
\hline
\endhead
\hline
\endfoot

GU Nor            & 1    & 3.45294  & 101         & 3.452940 & 2,3     \\
DW Per            & 1    & 3.649962 & 101         & 3.64998  & 3       \\
SS Sct & 1   & 3.67134  & H24 & 3.67140  & 1 \\
RT Aur & 1   & 3.728309 & H24 & 3.728324 & 4,5 \\
MU Cep            & 1    & 3.767843 & 101         & 3.76789  & 3       \\
SU Cyg & 1   & 3.845548 & H24 & 3.84557  & 2 \\
CS Ori            & 1a   & 3.889113 & 101        & 3.88913  & 2,3    \\
EK Mon            & 1    & 3.957979 & 101         & 3.957971 & 2,3     \\
ST Tau & 1a  & 4.034239 & H24 & 4.03436  & 1 \\
BF Oph & 1a  & 4.06754  & H24 & 4.067664 & 2 \\
SY Cas & 1   & 4.071161 & H24 & 4.07115  & 2 \\
Y Lac  & 1a  & 4.323775 & H24 & 4.32376  & 2 \\
V402 Cyg & 1a & 4.364926 & H24 & 4.36493  & 2 \\
V496 Cen          & 1    & 4.424150 & 101         & 4.424195 & 2,6,10    \\
T Vul & 1   & 4.435408 & H24 & 4.43542  & 2 \\
XY Cas & 1   & 4.5017   & H24 & 4.501692 & 2,3 \\
V407 Cas          & 1    & 4.566102 & 101         & 4.566090 & 2,3     \\
S Cru  & 1   & 4.68972  & H24 & 4.689731 & 2 \\
VZ Cyg & 1   & 4.864331 & H24 & 4.86440  & 2,8 \\
CF Cas & 1   & 4.87511  & H24 & 4.87511  & 2 \\
V1154 Cyg & 1 & 4.925468 & H24 & 4.92548 & 2 \\
AP Sgr & 1a  & 5.05806  & H24 & 5.057937 & 2 \\
V381 Cen          & 1    & 5.07884  & 101         & 5.07884  & 2       \\
AP Pup & 1   & 5.08453  & H24 & 5.08428  & 2 \\
V350 Sgr & 1 & 5.154239 & H24 & 5.15434  & 2 \\
V386 Cyg & 1 & 5.25762  & H24 & 5.25769  & 2 \\
AX Cir & 1a  & 5.27351  & H24 & 5.27340  & 2 \\
FN Vel            & 1    & 5.324191 & 101         & 5.32415  & 1,2     \\
BG Lac & 1   & 5.33194  & H24 & 5.33194  & 2,5 \\
UY Per            & 1a   & 5.365150 & 101         & 5.36512  & 3       \\
$\delta$ Cep & 1 & 5.366311 & H24 & 5.366263 & 2,4,5 \\
V1162 Aql & 1 & 5.37620 & H24 & 5.37619 & 2 \\
CV Mon & 1a  & 5.37866  & H24 & 5.37865  & 2 \\
V Cen             & 1    & 5.494067 & 101         & 5.49405  & 2       \\
VY Per            & 1a   & 5.531909 & 101         & 5.531919 & 2,3     \\
ASAS182714-1507.1 & 1a   & 5.545536 & 101         & 5.54582  & 1       \\
Y Sgr  & 1a  & 5.773380 & H24 & 5.77339 & 2 \\
FM Cas & 1   & 5.80931  & H24 & 5.8095   & 5 \\
AO CMa            & 1    & 5.815665 & 101         & 5.81564  & 2,3     \\
R Cru  & 1a  & 5.82591  & H24 & 5.82587  & 2 \\
T Ant & 1 & 5.898428  & 101 & 5.898393 & 1,2\\
MW Cyg & 1   & 5.95489  & H24 & 5.95466  & 2 \\
RV Sco            & 1    & 6.061332 & 101         & 6.06135  & 2       \\
FM Aql & 1   & 6.11411  & H24 & 6.11423  & 2 \\
V538 Cyg & 1 & 6.11906  & H24 & 6.11904  & 2,3,9 \\
V733 Aql & 1a & 6.17863 & H24 & 6.17859  & 2 \\
RS Nor            & 1a   & 6.198218 & 101,102,103 & 6.19823  & 1       \\
CR Cep & 1   & 6.2333   & H24 & 6.23345  & 3,9 \\
RR Lac & 1a  & 6.41629  & H24 & 6.41634  & 2 \\
XX Sgr & 1a  & 6.42425  & H24 & 6.42434  & 2 \\
AW Per & 1   & 6.463634 & H24 & 6.46376  & 5,7 \\
BB Sgr & 1   & 6.63712  & H24 & 6.63714  & 2 \\
CS Mon            & 1    & 6.73198  & 101         & 6.73196  & 1,2,3   \\
T Cru  & 1   & 6.7334   & H24 & 6.73307  & 2 \\
U Sgr  & 1   & 6.745310 & H24 & 6.74532  & 2 \\
V636 Sco & 1 & 6.796959 & H24 & 6.79695  & 2 \\
V496 Aql & 1 & 6.80707  & H24 & 6.80706  & 1 \\
V397 Nor          & 1    & 6.81262  & 101         & 6.81258  & 3,6     \\
BG Vel & 1   & 6.9239   & H24 & 6.92378  & 2 \\
X Sgr  & 1a  & 7.01272  & H24 & 7.01275  & 2 \\
U Aql  & 1   & 7.024149 & H24 & 7.02411  & 2 \\
V737 Cen          & 1    & 7.06591  & 101         & 7.06606  & 2       \\
TW Mon            & 1    & 7.096998 & 101,104     & 7.09693  & 2,3    \\
$\eta$ Aql & 1 & 7.17648 & H24 & 7.17688 & 2 \\
GX Car            & 1    & 7.196935 & 101         & 7.19684  & 1       \\
V600 Aql & 1 & 7.23880  & H24 & 7.23885  & 2 \\
V459 Cyg & 1 & 7.25146  & H24 & 7.25151  & 3 \\
TZ Mon & 1   & 7.42813  & H24 & 7.42825  & 1,2,3 \\
V1344 Aql & 1 & 7.47678 & H24 & 7.47682  & 1,2 \\
BB Her & 1   & 7.50794  & H24 & 7.50800  & 9 \\
R Mus             & 1    & 7.510331 & 101         & 7.51043  & 2       \\
IT Car & 1   & 7.540    & H24 & 7.53312  & 2 \\
RS Ori & 1a  & 7.56697  & H24 & 7.5670   & 1 \\
V492 Cyg          & 1    & 7.57801  & 101         & 7.5780   & 3       \\
W Sgr  & 1   & 7.595015 & H24 & 7.59501  & 2 \\
GH Cyg & 1   & 7.81796  & H24 & 7.81774  & 2 \\
VY Cyg & 1a  & 7.85723  & H24 & 7.85717  & 2 \\
RX Cam & 1   & 7.91220  & H24 & 7.91209  & 9 \\
W Gem  & 1a  & 7.91336  & H24 & 7.9131   & 1 \\
V2340 Cyg         & 1    & 7.96664  & 101         & 7.9666   & 3       \\
U Vul  & 1   & 7.990701 & H24 & 7.99079  & 2 \\

\hline
\end{longtable}
    \begin{tablenotes}
    \item \textbf{Notes:} \small The quality flags (2nd column) as defined in \cite{Hocde2024RV} are either excellent "1" or "1a" with slightly unstable fit. We list the pulsation period that we have derived for each star and which is used in the Fourier fit, see Sect.~\ref{sect:fourier} for details. \textbf{RV References:} 101: \cite{Anderson2024}, 102: \cite{Metzger1992}, 103: \cite{Pont1994},
104: \cite{Pont1997}, 105: \cite{Pont2001}. \textbf{LC References:} 1: \cite{Pojmanski2002}
 2: \cite{Berdnikov2008}
 3: ASASSN-V \cite{ASAS2018}
 4: \cite{kiss98}
 5: \cite{moffett84}
 6: \cite{Berdnikov2015}
 7: \cite{szabados80}
 8: \cite{barnes97}
9: KWS \citep{Morokuma2014}
10: AAVSO \citep{Kloppenborg2025}
    \end{tablenotes}

\begin{longtable}{l l | S[table-format=2.6] c | S[table-format=2.6] c}
\caption{Sample of metal-poor Cepheids  \label{tab:data_ref_metal_poor}}
 \\
\hline\hline
Name & Quality & \multicolumn{2}{c|}{RV curve} & \multicolumn{2}{c}{LC} \\
     &         & \multicolumn{1}{c}{Period} & Ref. & \multicolumn{1}{c}{Period} & Ref. \\
\hline
\endfirsthead
\hline\hline
Name & Quality & \multicolumn{2}{c|}{Radial velocity curve} & \multicolumn{2}{c}{LC} \\
     &         & \multicolumn{1}{c}{Period} & Ref. & \multicolumn{1}{c}{Period} & Ref. \\
\hline
\endhead
\hline
\endfoot
FI Mon & 2   & 3.287824 & H24 & 3.28776  & 3 \\
 FT Mon                  & 2   & 3.42177  & H24          & 3.421766  & 3,6\\
BC Pup            & 2   & 3.54422  & H24        & 3.54431  & 3     \\
FG Mon            & 2   & 4.496620 & H24        & 4.49664  & 3     \\
WW Mon            & 2   & 4.662169 & 101,104,105 & 4.66224  & 3    \\
CU Mon            & 2  & 4.70756  & H24        & 4.70766  & 3     \\
XX Mon            & 2 & 5.45654  & H24        & 5.45669  & 3     \\
V510 Mon          & 2  & 7.45750  & H24        & 7.4580   & 3     \\
\hline
OGLE-SMC-CEP-1761 &   2     & 3.93920  & 200         & 3.939278 & 21    \\
OGLE-SMC-CEP-1729 &    2    & 4.28430  & 200         & 4.284396 & 21    \\
OGLE-SMC-CEP-1680 &     2   & 4.88863  & 200         & 4.88837  & 20      \\
OGLE-SMC-CEP-1765 &      2  & 5.62408  & 200         & 5.62400  & 20      \\
OGLE-SMC-CEP-1717 &       2 & 6.66049  & 200         & 6.6605   & 20      \\
\hline
OGLE-LMC-CEP-3724 &   2     & 2.639172 & 201,204,206 & 2.63920  & 20      \\
OGLE-LMC-CEP-3750 &    2    & 2.95413  & 204,206     & 2.95413  & 20      \\
OGLE-LMC-CEP-4506 &  2      & 2.98788  & 202         & 2.98779  & 20      \\
OGLE-LMC-CEP-4646 &   2     & 3.10121  & 204,206     & 3.10118  & 20,22        \\
OGLE-LMC-CEP-3723 &   2     & 3.14376  & 201,204,206 & 3.14380  & 20      \\
OGLE-LMC-CEP-3726 &   2     & 3.52279  & 203,204,206 & 3.522852 & 20      \\
OGLE-LMC-CEP-227  &   2     & 3.79708  & 202         & 3.79706  & 20      \\
OGLE-LMC-CEP-3320 &   2     & 4.785034 & 205         & 4.78500  & 20      \\
OGLE-LMC-CEP-1249 &    2    & 6.878425 & 205         & 6.87867  & 20      \\
OGLE-LMC-CEP-1327 &    2    & 6.923806 & 205         & 6.92381  & 20      \\
\hline
\end{longtable}
    \begin{tablenotes}
    \item \textbf{Notes.} \small The quality flag "2" refers to RV curves of poor quality because either of unstable fit, insufficient coverage or small number of data points. We list the pulsation period that we have derived for each star and which is used in the Fourier fit, see Sect.~\ref{sect:fourier} for details. \textbf{RV~References:} 101: \cite{Anderson2024},
104: \cite{Pont1997}, 105: \cite{Pont2001}, 200: \cite{Gieren2018}, 201: \cite{Molinaro2012}, 202: \cite{Pilecki2018}, 203: \cite{Storm2004a}
 204: \cite{Storm2005}, 205: \cite{storm11b}, 206: \cite{Welch1991}. \textbf{LC References:} 
 3: ASASSN-V, \cite{ASAS2018}
6: \cite{Berdnikov2015}
20: OGLE IV, \cite{Soszynski2015}
21: OGLE III, \cite{Soszynski2008,Soszynski2010}
22: \cite{gieren2000}
    \end{tablenotes}

\begin{table*}
\centering
\caption{Fourier parameters of the $V$-band LCs for the calibrating sample.}\label{tab:fourier_lc}
\begin{tabular}{lccccccccccc}
\hline
Star & $P$ (day) & $n$ & Ndat & $\sigma_\mathrm{fit}$(mag) & $A_1$(mag) & $R_{21}$ & $\phi_{21}$ & $R_{31}$ & $\phi_{31}$ & $R_{41}$ & $\phi_{41}$ \\
\hline

GU Nor
& 3.45294 &     4    &229 &0.023 & 0.2296 & 0.311 & 2.603 & 0.110 & 5.243 & 0.041 & 1.291 \\
&         &        & && 0.0023 & 0.010 & 0.036 & 0.010 & 0.097 & 0.010 & 0.243 \\

DW Per
& 3.649978 & 6 &199 &0.013 & 0.2853 & 0.349 & 2.627 & 0.159 & 5.413 & 0.079 & 1.810 \\
&       &   &   &     & 0.0014 & 0.005 & 0.016 & 0.005 & 0.033 & 0.005 & 0.059 \\

SS Sct
& 3.671403 & 5 &187 &0.008 & 0.2301 & 0.304 & 2.632 & 0.117 & 5.300 & 0.040 & 1.630 \\
&       &   &&        & 0.0009 & 0.004 & 0.014 & 0.004 & 0.034 & 0.004 & 0.094 \\
...
&  &  && &  &  &  &  &  & & \\
&     &  &   &        &  & &  & &  &  &  \\
\hline
\end{tabular}
\begin{tablenotes}
\item \textbf{Notes.} This table presents for each star the pulsation period $P$, the order of the fit $n$, the RMS of the fit $\sigma_\mathrm{fit}$(mag), and the Fourier parameters up to fourth order with their uncertainties. The full table including all derived parameters and uncertainties is available electronically at CDS. 
\end{tablenotes}
\end{table*}

\begin{table*}
\centering
\caption{Fourier parameters of the RV curves for the calibrating sample.}\label{tab:fourier_rv}
\begin{tabular}{lccccccccccc}
\hline
Star & $P$ (day) & $n$ & Ndat & $\sigma_\mathrm{fit}$(km/s) & $A_1$(km/s) & $R_{21}$ & $\phi_{21}$ & $R_{31}$ & $\phi_{31}$ & $R_{41}$ & $\phi_{41}$ \\
\hline

GU Nor
& 3.452941 &  5 & 26 & 0.090 & 13.44 & 0.270 & 3.038 & 0.095 & 6.053 & 0.034 & 2.840 \\
&          &  &  &        & 0.03  & 0.003 & 0.012 & 0.004 & 0.022 & 0.003 & 0.067 \\

DW Per
& 3.649962 &  7 &24 & 0.100 & 14.90 & 0.325 & 2.882 & 0.126 & 5.815 & 0.063 & 2.511 \\
&          &   & &        & 0.04  & 0.003 & 0.009 & 0.003 & 0.029 & 0.004 & 0.036 \\

SS Sct
& 3.671338 &  3 & 67 & 0.780 & 14.18 & 0.268 & 2.917 & 0.095 & 5.933 & --- & --- \\
&          &   & &        & 0.14  & 0.010 & 0.041 & 0.010 & 0.108 & --- & --- \\
...
&  &  & & & &  &  &  &  & & \\
&       &  & &        &  & &  & &  &  &  \\
\hline
\end{tabular}
\begin{tablenotes}
\item \textbf{Notes.} Same as Table \ref{tab:fourier_lc}. The full table including all derived parameters and uncertainties is available electronically at CDS.
\end{tablenotes}
\end{table*}

\end{appendix}
\end{document}